\documentclass[10pt, conference]{IEEEtran}
\usepackage[noadjust]{cite}
\usepackage{multirow}
\usepackage{epsfig,endnotes}
\usepackage{subfig}
\usepackage{graphicx}
\usepackage{grffile}
\usepackage[font=bf]{caption}
\usepackage{color, url}
\usepackage{xspace} 
\usepackage{mathrsfs}
\usepackage{amssymb}
\usepackage{amsmath}
\usepackage{epstopdf}
\usepackage{balance}
\usepackage{bm}
\usepackage{rotating}
\let\labelindent\relax
\usepackage{enumitem}
\usepackage{makecell}
\setlist{nolistsep}

\usepackage{algorithm}
\usepackage{algorithmic}

\newtheorem{definition}{Definition}

\newcommand{\myparatight}[1]{\smallskip\noindent{\bf {#1}:}~}

\newcommand{\alan}{}

\graphicspath{ {../fig/} }

\begin{document}

\title{Graph-based Security and Privacy Analytics \\ via Collective Classification \\ with Joint Weight Learning and Propagation}

\author{
    \IEEEauthorblockN{Binghui Wang, Jinyuan Jia, Neil Zhenqiang Gong}
    \IEEEauthorblockA{Electrical and Computer Engineering Department, Duke University}
    \IEEEauthorblockA{\{binghui.wang, jinyuan.jia, neil.gong\}@duke.edu}
}

\IEEEoverridecommandlockouts
\makeatletter\def\@IEEEpubidpullup{6.5\baselineskip}\makeatother
\IEEEpubid{\parbox{\columnwidth}{
    Network and Distributed Systems Security (NDSS) Symposium 2019\\
    24-27 February 2019, San Diego, CA, USA\\
    ISBN 1-891562-55-X\\
    https://dx.doi.org/10.14722/ndss.2019.23226\\
    www.ndss-symposium.org
}
\hspace{\columnsep}\makebox[\columnwidth]{}}

\maketitle

\begin{abstract}
Many security and privacy problems can be modeled as a graph classification problem, where nodes in the graph are classified by \emph{collective classification} simultaneously. 
State-of-the-art collective classification methods for such graph-based security and privacy analytics follow the following paradigm: assign weights to edges of the graph, iteratively propagate \emph{reputation scores} of nodes among the weighted graph, and use the final reputation scores to classify nodes in the graph. 
The key challenge is to assign edge weights such that an edge has a large weight if the two corresponding nodes have the same label, and a small weight otherwise. 
Although collective classification has been studied and applied for security and privacy problems for more than a decade, how to address this challenge is still an open question. For instance, 
 most existing methods simply set a constant weight to all edges.

In this work, we propose a novel collective classification framework to address this long-standing challenge. We first formulate learning edge weights as an optimization problem, which quantifies the goals about the final reputation scores that we aim to achieve. However, it is computationally hard to solve the optimization problem because the final reputation scores depend on the edge weights in a very complex way. To address the computational challenge, we propose to jointly learn the edge weights and propagate the reputation scores, which is essentially an approximate solution to the optimization problem. We compare our framework with state-of-the-art methods for graph-based security and privacy analytics using four large-scale real-world datasets from various application scenarios such as Sybil detection in social networks, fake review detection in Yelp, and attribute inference attacks. Our results demonstrate that our framework achieves higher accuracies than state-of-the-art methods with an acceptable computational overhead.

\end{abstract}

\section{Introduction}
\label{intro}

Graphs are a powerful tool to represent complex interactions between various entities.
A particular family of graph-based machine learning techniques called \emph{collective classification}~\cite{sen2008collective} have been applied to various security and privacy problems, including malware detection~\cite{chau2011polonium,ye2011combining,rajab2013camp,tamersoy2014guilt,stringhini2017marmite}, 
 Sybil detection in social networks~\cite{Yu06, Yu08, Danezis09, Viswanath10, Mohaisen11, Yang11-sybil, sybilrank, integro, sybilbelief, robustspammer, wang2017sybilscar, wang2018structure,jia2017random,wang2017gang,zheng2018smoke,gao2018sybilfuse}, 
fake review detection~\cite{akoglu2013opinion, li2014spotting, rayana2015collective}, 
 malicious website detection~\cite{gyongyi2004combating, wu2006propagating,li2013finding}, auction fraud detection~\cite{pandit2007netprobe}, APT infection detection~\cite{Oprea15},
and attribute inference attacks~\cite{Zheleva09, gong2014joint, Chaabane12, GongAttributeInfer16, jia2017attriinfer, olteanu2017quantifying, gong2018attribute}. Moreover,  some collective classification methods have been deployed in industry, e.g., Symantec deployed collective classification to detect malware~\cite{tamersoy2014guilt} and  Tuenti, the largest social network in Spain, deployed collective classification to detect Sybils~\cite{sybilrank,integro}.

\begin{figure}[!t]
\center
{\includegraphics[width=0.35\textwidth]{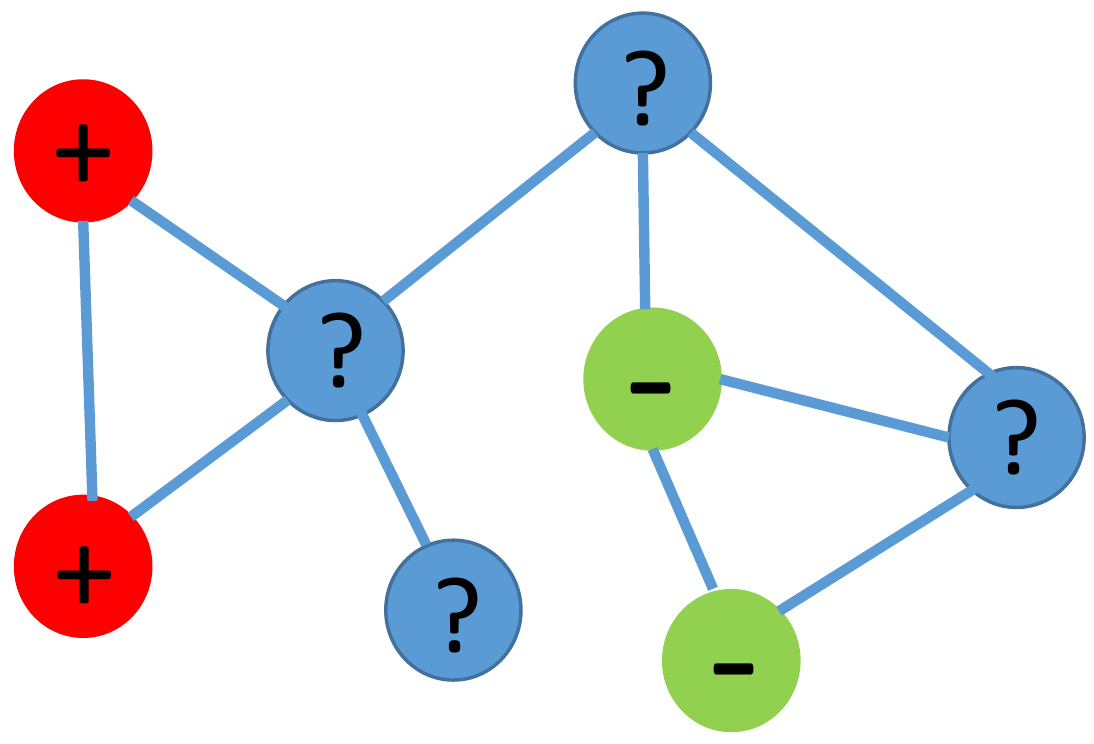} }
\caption{Illustration of collective classification.}
\vspace{-6mm}
\label{example}
\end{figure}

Figure~\ref{example} illustrates the setting of collective classification: given 1) a graph, which can be either undirected or directed, and 2) a training dataset, which consists of some labeled positive nodes and/or labeled negative nodes, collective classification is to classify the remaining unlabeled nodes to be positive or negative \emph{simultaneously}. 
For different security and privacy applications, nodes represent different entities, edges represent different relationships, and the labels ``positive" and ``negative" have different semantic meanings. Table~\ref{security_privacy_problems} shows the possible meanings of nodes, edges, positive, and negative in several security and privacy problems. We will discuss more details about how these problems were modeled as collective classification in Section~\ref{relatedwork}.

\begin{table*}[!t]\renewcommand{\arraystretch}{1.8}
\centering
\caption{Possible meanings of nodes, edges, positive, and negative in example security and privacy application scenarios.}
\centering
\begin{tabular}{|c|c|c|c|c|} \hline 
{\normalsize \bf Security \& Privacy Problem} 	& {\normalsize \bf Nodes} & {\normalsize \bf Edges} &{\normalsize \bf Positive} &{\normalsize \bf Negative} \\ \hline
{\small Sybil Detection} &  {\small Users} & {\small Friendship, Following} & {\small Sybil} & {\small Benign User}  \\ \hline
{\small Fake Review Detection} &  {\small Users, Reviews, Products} & {\small Reviewing} & {\small Fake Review} & {\small Genuine Review}  \\ \hline
{\small Malware Detection} &  {\small Files, Hosts} & {\small Appearance} & {\small Malware} & {\small Benign File}  \\ \hline
{\small Malicious Website Detection} &  {\small Websites} & {\small Redirection} & {\small Malicious Website} & {\small Benign Website}  \\ \hline
{\small Attribute Inference Attack} & {\small Users} & {\small Friendship} & {\small Having the Attribute} & {\small Not Having the Attribute}\\ \hline
\end{tabular}
\label{security_privacy_problems}
\end{table*}

State-of-the-art collective classification methods~\cite{gyongyi2004combating, wu2006propagating,li2013finding,pandit2007netprobe,chau2011polonium,Mohaisen11,sybilrank,sybilbelief,tamersoy2014guilt,akoglu2013opinion, li2014spotting, rayana2015collective,stringhini2017marmite,wang2017sybilscar, jia2017random, wang2017gang,integro,wang2018structure,gao2018sybilfuse} 
 have three key steps. In Step I, they assign a \emph{prior reputation score} to every node in the graph based on the training dataset. In Step II, they assign weights to edges in the graph, where a large weight of an edge $(u,v)$ indicates that $u$ and $v$ are likely to have the same label. 
In Step III, they iteratively propagate the prior reputation scores among the weighted graph to obtain a \emph{posterior reputation score} for every node. The posterior reputation scores are used to classify or rank nodes. The details of one or more of the three steps could be different for different collective classification methods. For instance, some methods leverage {random walks} for propagation in Step III, while some methods leverage loopy belief propagation in Step III, which we call \emph{random walk (RW)-based methods} and \emph{loopy belief propagation (LBP)-based methods}, respectively. 

An edge $(u,v)$ is said to be \emph{homogeneous} if $u$ and $v$ have the same label, otherwise it is said to be \emph{heterogeneous}. The propagation in Step III requires that homogeneous edges have large  weights and heterogeneous edges have small weights. 
However, existing methods face a key limitation: they assign small weights to a large number of {homogeneous} edges and/or large weights to a large number of {heterogeneous} edges in Step II. 
Specifically, most existing methods~\cite{pandit2007netprobe,chau2011polonium,ye2011combining,sybilrank,sybilbelief,tamersoy2014guilt,akoglu2013opinion, li2014spotting, rayana2015collective,wang2017sybilscar,wang2018structure, jia2017random, wang2017gang} simply set a large constant weight to all edges, ignoring the difference between homogeneous and heterogeneous edges. A few methods~\cite{integro,gao2018sybilfuse} proposed to learn edge weights using features extracted from nodes' characteristics and the graph structure. For instance, to detect Sybils in social networks, \'{I}ntegro~\cite{integro} first learns a classifier based on users' profiles to predict the probability that a node is a victim (a victim is a benign user connecting to at least one Sybil) and then assigns small weights to all edges of the nodes that are victims with high probabilities. However, \'{I}ntegro assigns small weights to  a large number of homogeneous edges, because a large number of nodes are victims~\cite{gao2018sybilfuse}. 
SybilFuse~\cite{gao2018sybilfuse} extracts features from the graph structure for each node, learns a classifier to predict the prior reputation score for each node using the training dataset, and uses the prior reputation scores to assign edge weights. In particular, an edge has a larger weight if the two nodes of the edge are more likely to have the same label, where a node's label is determined by its prior reputation score.  
Since the prior reputation scores are inaccurate at determining nodes' labels (otherwise we do not need to propagate the prior reputation scores to obtain posterior reputation scores in Step III), a large number of heterogeneous edges are assigned large weights while a large number of homogeneous edges are assigned small weights. 

As a result, existing methods have a limited success in the security and privacy problems that have a large amount of heterogeneous edges, e.g., Sybil detection in weak-trust social networks like Twitter, fake review detection, and attribute inference.

In this work, we propose a new framework to learn edge weights for graph-based security and privacy analytics. Our framework is applicable to both RW-based and LBP-based methods. 
Our key intuition is that the edge weights and the final posterior reputation scores produced by a collective classification method should satisfy two goals. First, the labeled positive nodes and the labeled negative nodes in the training dataset should have high and low posterior reputation scores, respectively. Second, the edge weights and the posterior reputation scores should be \emph{consistent}. Specifically, we use the final posterior reputation scores to predict labels of nodes; an edge is predicted to be homogeneous if the two nodes of the edge are predicted to have the same label, otherwise the edge is predicted to be heterogeneous. Consistency between edge weights and posterior reputation scores means that an edge that is predicted to be homogeneous should have a large weight, while an edge that is predicted to be heterogeneous should have a small weight.

We formulate learning edge weights as minimizing an \emph{objective function}, where the objective function is small if the two goals are achieved. However, it is computationally challenging to solve the formulated optimization problem (we will discuss more details in Section~\ref{optimizationproblem}) because the final posterior reputation scores depend on the edge weights in a very complex way, e.g., every edge weight could influence the final posterior reputation score of every node.  

To address the computational challenge, we propose to jointly learn the edge weights and propagate the posterior reputation scores, which can be viewed as an approximate solution to our formulated optimization problem. 
Our key idea is that the posterior reputation scores are iteratively updated in Step III and thus we iteratively learn edge weights using the current posterior reputation scores instead of the final posterior reputation scores. 
Specifically,  
given the posterior reputation scores in the $t$th iteration, we learn new edge weights and then use the learnt edge weights to update the posterior reputation scores in the $(t+1)$th iteration. We aim to learn the edge weights to satisfy the two goals: 1) the posterior reputation scores in  the $(t+1)$th iteration should be large and small for the labeled positive nodes and negative nodes in the training dataset, respectively; and 2) the learnt edge weights should be consistent with the posterior reputation scores in the $t$th iteration. Learning edge weights is efficient under our framework because the posterior reputation scores of the nodes in the training dataset in the $(t+1)$th iteration only depend on weights of the edges of the nodes in the training dataset. 

We compare our framework with state-of-the-art RW-based and LBP-based methods using multiple large-scale real-world datasets, which are from different application scenarios including Sybil detection in social networks, fake review detection, and attribute inference attacks.  For instance, the Twitter dataset for Sybil detection has 42M users and 1.5B edges, while the Google+ dataset for attribute inference attacks has 5M users and 31M edges. 
Our results demonstrate that our framework has significantly higher accuracies than state-of-the-art methods, with an acceptable computational overhead (around 2-3 times slower than state-of-the-art methods).    
Moreover, we apply our framework to detect Sybils in a large-scale directed Sina Weibo (a large social network in China) dataset with 3.5M users and 653M edges.  
We manually verify the detected Sybils and 
our results show that our framework can accurately detect Sybils, e.g., 95\% of the top-ranked 100K users are Sybils.

We summarize our contributions as follows:
\begin{itemize}
\item We propose a novel framework to learn edge weights for graph-based security and privacy analytics. 
\item We formulate learning edge weights as solving an optimization problem. Moreover, we design efficient algorithms to solve the optimization problem. 
\item We compare our framework with state-of-the-art methods using multiple large-scale real-world datasets from different security and privacy application scenarios.  
\end{itemize}

\begin{table*}[!t]\renewcommand{\arraystretch}{1.8}
\centering
\caption{Representative methods developed for different security and privacy applications. 
\emph{Training Dataset}: Negative, Positive, or Both means that  a method leverages labeled negative nodes, labeled positive nodes, or both to assign prior reputation scores. 
\emph{Weight Learning}: Y or N means that a method learns edge weights or does not. We note that all these methods assign edge weights and fix them in Step II. }
\centering
\begin{normalsize}
\begin{tabular}{|c|c|c|c|c|c|c|} \hline 
\multicolumn{2}{|c|}{\bf Method} 	& {\bf Application} & {\bf \makecell{Graph \\ Type}} &{\bf \makecell{Training \\ Dataset}} &{\bf \makecell{Weight \\ Learning}} &{\bf Publication} \\ \hline
\multirow{16}{*}{\bf RW} & {\bf SybilGuard~\cite{Yu06}} &  {\small Sybil Detection} & {\small Undirected} & {\small Negative}  & {\small N} & {\small SIGCOMM'06} \\ \cline{2-7}
{}& {\bf SybilLimit~\cite{Yu08}} &  {\small Sybil Detection} & {\small Undirected} & {\small Negative}  & {\small N} & {\small S\&P'08} \\ \cline{2-7}
{}& {\bf SybilInfer~\cite{Danezis09}} &  {\small Sybil Detection} & {\small Undirected} & {\small Negative}  & {\small N} & {\small NDSS'09} \\ \cline{2-7}
{}& {\bf SybilRank~\cite{sybilrank}} &  {\small Sybil Detection} & {\small Undirected} & {\small Negative} & {\small N} & {\small NSDI'12} \\ \cline{2-7}
{}& {\bf CIA~\cite{Yang12-spam}} &  {\small Sybil Detection} & {\small Directed} & {\small Positive}  & {\small N} & {\small WWW'12} \\ \cline{2-7}
{}& {\bf \'{I}ntegro~\cite{integro}} &  {\small Sybil Detection} & {\small Undirected} & {\small Negative} & {\small Y} & {\small NDSS'15} \\ \cline{2-7}
{\bf }& {\bf SmartWalk~\cite{LiuCCS16}} &  {\small Sybil Detection} & {\small Undirected} & {\small Negative}  & {\small N} & {\small CCS'16} \\ \cline{2-7}
{\bf }& {\bf SybilWalk~\cite{jia2017random}} &  {\small Sybil Detection} & {\small Undirected} & {\small Both} & {\small N} & {\small DSN'17} \\ \cline{2-7}
{\bf }& {\bf ELSIEDET~\cite{zheng2018smoke}} &  {\small Sybil Detection} & {\small Undirected} & {\small Positive} & {\small N} & {\small NDSS'18} \\ \cline{2-7}
{\bf }& {\bf TrustRank~\cite{gyongyi2004combating}} &  {\small Malicious Website Detection} & {\small Directed} & {\small Negative}  & {\small N} & {\small VLDB'04} \\ \cline{2-7}
{}& {\bf DistrustRank~\cite{wu2006propagating}} &  {\small Malicious Website Detection} & {\small Directed} & {\small Positive}  & {\small N} & {\small MTW'06} \\ \cline{2-7}
{}& {\bf Li et al.~\cite{li2013finding}} &  {\small Malicious Website Detection} & {\small Directed} & {\small Negative} & {\small N} & {\small S\&P'13} \\ \cline{2-7}
{}& {\bf Ye et al.~\cite{ye2011combining}} &  {\small Malware Detection} & {\small Undirected} & {\small Both}  & {\small N} & {\small KDD'11} \\ \cline{2-7}
{}& {\bf Marmite~\cite{stringhini2017marmite}} &  {\small Malware Detection} & {\small Undirected} & {\small Both} & {\small N} & {\small ACSAC'17} \\ \cline{2-7}
{}& {\bf RWwR-SAN~\cite{gong2014joint}} &  {\small Attribute Inference} & {\small Undirected} & {\small Positive}  & {\small N} & {\small TIST'14} \\ \cline{2-7}
{}& {\bf VAIL~\cite{GongAttributeInfer16}} &  {\small Attribute Inference} & {\small Undirected} & {\small Positive}  & {\small N} & {\small Security'16} \\ \cline{2-7}
{}& {\bf EdgeExplain~\cite{chakrabarti2017joint}} &  {\small Attribute Inference} & {\small Undirected} & {\small Both}  & {\small N} & {\small JMLR'17} \\ \hline
\multirow{11}{*}{\bf LBP}& {\bf SybilBelief~\cite{sybilbelief}} &  {\small Sybil Detection} & {\small Undirected} & {\small Both}  & {\small N} & {\small TIFS'14} \\ \cline{2-7}
{}& {\bf SybilSCAR~\cite{wang2017sybilscar,wang2018structure}} &  {\small Sybil Detection} & {\small Undirected} & {\small Both}  & {\small N} & {\small INFOCOM'17} \\ \cline{2-7}
{}& {\bf GANG~\cite{wang2017gang}} &  {\small Sybil Detection} & {\small Directed} & {\small Both} & {\small N} & {\small ICDM'17} \\ \cline{2-7}
{}& {\bf SybilFuse~\cite{gao2018sybilfuse}} &  {\small Sybil Detection} & {\small Undirected} & {\small Both} & {\small Y} & {\small CNS'18} \\ \cline{2-7}
{\bf }& {\bf Burst~\cite{fei2013exploiting}} &  {\small Fake Review Detection} & {\small Undirected} & {\small Both} & {\small N} & {\small ICWSM'13} \\ \cline{2-7}
{\bf }& {\bf FraudEagle~\cite{akoglu2013opinion}} &  {\small Fake Review Detection} & {\small Undirected} & {\small Both} & {\small N} & {\small ICWSM'13} \\ \cline{2-7}
{\bf }& {\bf SpEagle~\cite{rayana2015collective}} &  {\small Fake Review Detection} & {\small Undirected} & {\small Both} & {\small N} & {\small KDD'15} \\ \cline{2-7}
{}& {\bf Polonium~\cite{chau2011polonium}} &  {\small Malware Detection} & {\small Undirected} & {\small Both} & {\small N}  & {\small SDM'11} \\ \cline{2-7}
{}& {\bf AESOP~\cite{tamersoy2014guilt}} &  {\small Malware Detection} & {\small Undirected} & {\small Both} & {\small N} & {\small KDD'14} \\ \cline{2-7}
{}& {\bf NetProbe~\cite{pandit2007netprobe}} &  {\small Auction Fraud Detection} & {\small Undirected} & {\small Both}  & {\small N} & {\small WWW'07} \\ \cline{2-7}
{}& {\bf Olteanu et al.~\cite{olteanu2017quantifying}} &  {\small Attribute Inference} & {\small Undirected} & {\small Both}  & {\small N} & {\small TMC'17} \\  \cline{2-7}
{}& {\bf AttriInfer~\cite{jia2017attriinfer}} &  {\small Attribute Inference} & {\small Undirected} & {\small Both}  & {\small N} & {\small WWW'17} \\ \hline
\end{tabular}
\end{normalsize}
\label{graph_methods}
\end{table*}

\section{Related Work}
\label{relatedwork}

\myparatight{Modeling security/privacy problems as graph-based collective classification problems} Studies from multiple communities--such as security, data mining, and networking--have shown that various security and privacy problems can be modeled as collective classification problems on graphs. Specifically, given a graph (undirected or directed) and a training dataset consisting of labeled positive nodes and negative nodes, collective classification aims to classify the remaining unlabeled nodes simultaneously. 

For instance, in Sybil detection on social networks, the graph is a social graph, where nodes are users and edges represent friendships or following relationships between users; a positive node means a Sybil and a negative node means a benign user. In malware detection, a host-file graph is constructed to represent the relationships between hosts and executable files~\cite{chau2011polonium,ye2011combining,rajab2013camp,tamersoy2014guilt,stringhini2017marmite}. Specifically, in this graph, a node is a host or an executable file; an edge between a host and an executable file means that the file appears on the host; a positive node means a malware or compromised host, while a negative node means a benign file or normal host. In fake review detection, a user--review--product graph is constructed to represent the relationships between users, reviews, and products~\cite{rayana2015collective}. In this graph, a node is a user, a review, or a product; an edge between a user and a review means that the user writes the review; an edge between a review and a product means that the review is for the product; a positive node means a fake review while a negative node means a genuine review. In attribute inference attacks~\cite{Zheleva09, gong2014joint, GongAttributeInfer16, jia2017attriinfer}, the graph is the social graph between users; a node is positive if the node has a given attribute (e.g., republican), otherwise the node is negative. Table~\ref{security_privacy_problems} summarizes several security and privacy problems that can be modeled as collective classification on graphs. 
We note that some security problems~\cite{kwon2015dropper,Feng16,Xu17} can be modeled as graph analytics other than collective classification. However, we will focus on collective classification based security/privacy analytics.

State-of-the-art collective classification methods have three key steps: a \emph{prior reputation score} is assigned to every node in the graph based on the training dataset in Step I, a weight is assigned to every edge in the graph in Step II, and the prior reputation scores are propagated among the weighted graph to obtain a \emph{posterior reputation score} for every node in Step III. A larger reputation score means that the node is more likely to be positive. Different collective classification methods have different details with respect to the three steps.  For example, random walk (RW)-based methods use various random walks in Step III, while  loopy belief propagation (LBP)-based methods rely on LBP in Step III. Table~\ref{graph_methods} summarizes representative collective classification methods developed for different security and privacy applications. We note that a method developed for one application (e.g., Sybil detection) could also be applied to other applications (e.g., attribute inference). Next, we discuss RW-based methods and LBP-based methods. In particular, we will discuss the differences in their three steps.

\myparatight{RW-based methods} Depending on which labeled nodes in the training dataset are used to assign the prior reputation scores in Step I, we classify RW-based methods into three categories, i.e., \emph{RW-N (Negative)},  \emph{RW-P (Positive)}, and \emph{RW-B (Both)}.  RW-N~\cite{wu2006propagating,gyongyi2004combating,sybilrank,li2013finding,integro,GongAttributeInfer16} leverages labeled negative nodes to assign prior reputation scores, e.g., every labeled negative node in the training dataset has a prior reputation score of -1 (the absolute value of the reputation score does not matter since these RW-based methods rely on relative reputation scores) and the remaining nodes have prior reputation scores of 0. RW-P~\cite{wu2006propagating,Yang12-spam} leverages labeled positive nodes to assign prior reputation scores, e.g., every labeled positive node has a prior reputation score of 1 and the remaining nodes have prior reputation scores of 0. RW-B~\cite{zhu2003semi,ye2011combining,jia2017random}  leverages both labeled positive nodes and labeled negative nodes to assign prior reputation scores, e.g., labeled positive nodes, labeled negative nodes,  and unlabeled nodes have  prior reputation scores of 1, -1, and 0, respectively.

In Step II, most RW-based methods simply set a constant weight to all edges. \'{I}ntegro~\cite{integro}, developed under the context of Sybil detection in social networks, learns edge weights using features extracted from nodes. Specifically,  \'{I}ntegro first trains a node classifier based on nodes' features to predict the probability that a node is a victim (a victim is a negative node connecting to at least one positive node). Then, \'{I}ntegro assigns small weights to all edges of the nodes that are predicted to be victims. To be convenient with the notations, we denote \'{I}ntegro as \emph{RW-FLW (Fixed Learnt Weights)} since it learns the edge weights and fixes them in Step II. 

In Step III, different methods use different random walks to propagate the reputation scores. For instance, on an undirected graph, RW-N, RW-P, and RW-FLW iteratively distribute a node's current reputation score to its neighbors and a node sums the reputation scores distributed from its neighbors as its new reputation score. In particular, a node $u$ shares a fraction of its reputation score to a neighbor $v$, where the fraction is the edge weight divided by the weighted degree of $u$.  
 On a directed graph, RW-N~\cite{gyongyi2004combating} (or RW-P~\cite{wu2006propagating,Yang11-sybil}) shares a fraction of node $u$'s current reputation score to each outgoing (or incoming) neighbor, where the fraction is the edge weight divided by the total weights of $u$'s outgoing (or incoming) edges. 
 RW-B is applicable to undirected graphs and uses a different random walk to propagate reputation scores. Specifically, the fraction is the edge weight divided by the weighted degree of $v$ instead of $u$ when $u$ shares its reputation score to a neighbor $v$.

\myparatight{LBP-based methods}  These methods~\cite{pandit2007netprobe,chau2011polonium,sybilbelief,tamersoy2014guilt,akoglu2013opinion,li2014spotting,rayana2015collective,wang2017sybilscar,wang2017gang,gao2018sybilfuse,wang2018structure} leverage both labeled positive nodes and labeled negative nodes in the training dataset to assign prior reputation scores in Step I, e.g., labeled positive nodes, labeled negative nodes, and unlabeled nodes have prior reputation scores of 1, -1, and 0, respectively. 
 In Step II, most LBP-based methods simply set a constant weight to all edges in the graph. SybilFuse~\cite{gao2018sybilfuse}, developed under the context of Sybil detection, learns edge weights using features extracted from the graph structure. Specifically, SybilFuse first trains a node classifier using structure-based node features (e.g., local clustering coefficient) to predict the prior reputation score for each node. Then, SybilFuse uses the prior reputation scores to assign edge weights in a way that a larger edge weight is assigned if the two corresponding nodes are more likely to have the same label based on the prior reputation scores. 
We note that SybilFuse can also extract structure-based edge features (e.g., common neighbors between two nodes) to train an edge classifier to assign the edge weights. However, such edge weights are not effective in real-world graphs as shown in~\cite{gao2018sybilfuse}.  To stress the fact that SybilFuse learns edge weights and fixes them in Step II, we denote it as \emph{LBP-FLW}.

 In Step III, these LBP-based methods~\cite{pandit2007netprobe,chau2011polonium,sybilbelief,tamersoy2014guilt,akoglu2013opinion,li2014spotting,rayana2015collective} leverage either a standard LBP~\cite{Pearl88} or an optimized LBP~\cite{gatterbauer2015linearized,wang2017sybilscar,wang2017gang,GatterbauerLinLBP17,wang2018structure} to propagate the reputation scores among the weighted graph. 
The optimized LBP is an order of magnitude more efficient than the standard LBP. 
On an undirected graph, the optimized LBP shares a fraction of a node $u$'s reputation score to its neighbor $v$, where the fraction is simply the edge weight, and a node sums its prior reputation score and the reputation scores shared from its neighbors as its new reputation score in each iteration. On a directed graph,  the optimized LBP shares a node $u$'s reputation score to its neighbor $v$ like on an undirected graph if 
both the directed edges $(u,v)$ and $(v,u)$ exist. 
However, when $u$ connects with $v$ with only an outgoing edge, the optimized LBP  shares $u$'s reputation score with $v$ like on an undirected graph only if $u$'s current reputation score is negative. When $u$ connects with $v$ with only an incoming edge, the optimized LBP  shares $u$'s reputation score with $v$ only if $u$'s current reputation score is positive.

\section{Problem Definition}

Suppose we are given a graph (either undirected or directed) $G=(V,E)$, where $u \in V$ is a node and $(u,v) \in E$ is an edge, which indicates a certain relationship between $u$ and $v$. $|V|$ and $|E|$ are the number of nodes and edges, respectively. 
We are also given a training dataset $L$, which consists of a set of labeled positive nodes $L_P$ and a set of labeled negative nodes $L_N$. We denote by $y_u$ the label of node $u$, where $y_u=1$ means that $u$ is positive and $y_u=-1$ means that $u$ is negative. We formally define the collective classification  problem as follows:

\begin{definition}[Collective Classification Problem]
Suppose we are given 1) a graph (either undirected or directed) and 2) a training dataset including both labeled positive nodes and labeled negative nodes. Collective classification is to classify or rank the unlabeled nodes in the graph. 
\end{definition}

\subsection{Evaluation Metrics}
Like previous studies~\cite{Viswanath10, Mohaisen11, chau2011polonium,sybilrank,integro,sybilbelief,tamersoy2014guilt,wang2017sybilscar, wang2017gang}, we adopt the following metrics to evaluate a collective classification method for graph-based security and privacy analytics. 

\myparatight{Accuracy} One way to measure the accuracy of a collective classification method is to use its posterior reputation scores to classify the unlabeled nodes, i.e., a node is predicted to be positive if its posterior reputation score is bigger than a certain \emph{threshold}, and negative otherwise. Then, 
 accuracy is the fraction of unlabeled nodes whose labels are correctly predicted. However, such accuracy highly relies on the threshold. Moreover,  it is challenging  to select a good threshold for various collective classification methods. Therefore,   \emph{Area Under the Receiver Operating Characteristic Curve (AUC)} is widely used to measure the accuracy of a collective classification method.  AUC  does not depend on a threshold and can be consistently used to evaluate all collective classification methods that produce posterior reputation scores. In particular, we rank all unlabeled nodes in a decreasing order with respect to their posterior reputation scores. AUC  is the probability that a randomly selected positive node ranks higher than a randomly selected negative node. If a method ranks all positive nodes before negative nodes, then the method has an AUC of 1 (there exists a threshold such that all nodes are correctly classified); if  a method ranks all negative nodes before positive nodes, then the method has an AUC of 0 (there exists a threshold such that all nodes are incorrectly classified); a method has an AUC of 0.5 if the method ranks the unlabeled nodes uniformly at random.

\myparatight{Scalability} In real-world security and privacy problems, the graph often has a very large scale, e.g., hundreds of millions of nodes and edges. Therefore, a collective classification method should be scalable. In particular, we use the computational \emph{time} a method consumes to measure its scalability/efficiency.

\myparatight{Robustness to label noise} The training dataset is often obtained via manual inspection of nodes, or is crowdsourced from users or workers in a crowdsourcing platform such as Amazon Mechanical Turk~\cite{Wang13}. For instance, online social networks often provide users an option to report a certain user as a spammer or fraudulent user, which could be incorporated into the training dataset. However, due to human mistakes~\cite{Wang13,Freeman17}, the training dataset could have noise, i.e., some negative nodes are falsely labeled as positive and/or some positive nodes are falsely labeled as negative. Therefore, we also use robustness to label noise in the training dataset to evaluate collective classification methods. In particular, suppose $\alpha$\% of the nodes in the training dataset are falsely labeled and a method achieves an AUC of $\beta$ with such training dataset. Then, we say the method has a robustness of $\beta$ when the label noise is at the level of $\alpha$\%.

\section{Our Framework}
\label{model}

We first summarize state-of-the-art collective classification methods. In particular, the posterior reputation scores are essentially solutions to a \emph{system of equations}.
Then, we formulate learning edge weights as an optimization problem, which quantifies our goals on the posterior reputation scores. However, it is computationally challenging to solve this optimization problem because the posterior reputation scores depend on the edge weights in very complex ways. Finally, we introduce our strategy to alternately learn edge weights and propagate posterior reputation scores. Our strategy is an approximate solution to our formulated optimization problem.

\subsection{Background}
In state-of-the-art collective classification methods~\cite{gyongyi2004combating, wu2006propagating,li2013finding,pandit2007netprobe,chau2011polonium,Mohaisen11,sybilrank,Yang12-spam,sybilbelief,tamersoy2014guilt,akoglu2013opinion, li2014spotting, rayana2015collective,wang2017sybilscar, jia2017attriinfer,jia2017random, wang2017gang,integro,gao2018sybilfuse}, the posterior reputation scores are solutions to the following system of equations:
{
\begin{align}
\label{unified_linsys}
{\mathbf{p}} = f(\mathbf{q}, \mathbf{W}, \mathbf{p}), 
\end{align}
}%
where the column vector $\mathbf{q}$ is the nodes' prior reputation scores,  the column vector $\mathbf{p}$ is the nodes' posterior reputation scores, and the matrix $\mathbf{W}$ is the edge weight matrix. Different methods have different prior reputation scores $\mathbf{q}$, assign different edge weight matrix $\mathbf{W}$, and use different function $f$. 
Next, we discuss the choices for these parameters in state-of-the-art methods.  Since LBP-based methods outperform RW-based methods~\cite{koutra2011unifying,jia2017attriinfer,wang2017sybilscar,wang2017gang}, we will focus on LBP-based methods for simplicity. However, we stress that our framework is also applicable to RW-based methods (see more details in Section~\ref{discussion}). 

\myparatight{LBP-based methods on undirected graphs} These methods~\cite{pandit2007netprobe,chau2011polonium,sybilbelief,tamersoy2014guilt,jia2017attriinfer,wang2017sybilscar,gao2018sybilfuse} often assign a positive prior reputation score to a labeled positive node and a negative prior reputation score to a labeled negative node in the training dataset. For example, the prior reputation score ${q}_u$ of the node $u$ can be assigned as follows:
{
 \begin{align}
 \label{prior}
 {q}_u=
 \begin{cases} 
\theta  & \text{if } u \in L_P   \\ 
-\theta  &  \text{if } u \in L_N \\
0 & \text{otherwise},
\end{cases} 
 \end{align}
}
where $\theta >0$ (e.g., $\theta=1$ in our experiments), $L_P$ is the set of labeled positive nodes, and $L_N$ is the set of labeled negative nodes in the training dataset. The entry $w_{uv}$ of the matrix $\mathbf{W}$ is 0 if $u$ and $v$ are not connected, otherwise $w_{uv}$ and $w_{vu}$ are the weight of the edge $(u,v)$. The edge weight $w_{uv}$ indicates the likelihood that $u$ and $v$ have the same label. Specifically, $w_{uv} > 0$ means that $u$ and $v$ are likely to have the same label, i.e., the edge  $(u,v)$ is \emph{homogeneous}; $w_{uv} < 0$ means that $u$ and $v$ are likely to have different labels, i.e., the edge $(u,v)$ is \emph{heterogeneous}; $w_{uv} = 0$ means that $u$ and $v$ are not correlated. 
Moreover, the function $f$ is as follows:
{
\begin{align}
\label{sybilscar}
 f(\mathbf{q}, \mathbf{W}, \mathbf{p}) = {\mathbf{q}} + {\mathbf{W}} {\mathbf{p}},
\end{align}
}%
where we consider the optimized LBP~\cite{gatterbauer2015linearized,wang2017sybilscar, wang2017gang,GatterbauerLinLBP17} instead of the standard LBP~\cite{Pearl88}. 

\myparatight{LBP-based methods on directed graphs} These methods~\cite{wang2017gang} also assign the prior reputation scores in Equation~\ref{prior}. Moreover, each connected node pair $(u,v)$ has two entries $w_{uv}$ and $w_{vu}$ in the weight matrix $\mathbf{W}$, where $w_{uv}$ does not necessarily equal $w_{vu}$. However, they use a different function $f$ as follows:  
{\small
\begin{align}
\label{gang}
f(\mathbf{q}, \mathbf{W}, \mathbf{p}) = {\mathbf{q}} + (\mathbf{W} \circ \mathbf{A}_b) {\mathbf{p}} + (\mathbf{W} \circ \mathbf{A}_i) \mathcal{I}(\mathbf{p}) + (\mathbf{W} \circ \mathbf{A}_o) \mathcal{J}(\mathbf{p}),
\end{align}
}%
where the operator $\circ$ represents  element-wise product of two matrices; the matrix $A_b$ is the adjacency matrix for bidirectional edges, i.e., if both the edge $(u,v)$ and the edge $(v,u)$ exist, then $A_{b,uv}=A_{b,vu}=1$, otherwise $A_{b,uv}=A_{b,vu}=0$; $A_i$ is the adjacency matrix for unidirectional incoming edges, i.e., if  $(v,u)$ exists but $(u,v)$ does not, then $A_{i,uv}=1$, otherwise $A_{i,uv}=0$; $A_o$ is the adjacency matrix for unidirectional outgoing edges, i.e., if  $(u,v)$ exists but $(v,u)$ does not, then $A_{o,uv}=1$, otherwise $A_{o,uv}=0$;  $\mathcal{I}$ and $\mathcal{J}$ are functions that apply to every entry of $\mathbf{p}$, and they reset the positive and negative entries to be 0, respectively.

Suppose $v$ is a neighbor of $u$. In LBP-based methods, a bidirectional neighbor $v$ multiplies its posterior reputation score by the edge weight and shares it with the node $u$; an incoming neighbor $v$ does so only if $v$'s posterior reputation score is negative; and an outgoing neighbor $v$ does so only if $v$'s posterior reputation score is positive. The intuition is that an incoming neighbor $v$  influences the node $u$'s label only if $v$ is predicted to be negative, while an outgoing neighbor $v$ influences the node $u$'s label only if $v$ is predicted to be positive. For instance, in Sybil detection on Twitter, if $v$ follows $u$ but $u$ does not follow back, then $v$ is informative for determining $u$'s label only if $v$ is benign, because a Sybil can arbitrarily follow any users; if $u$ follows $v$ but $v$ does not follow back, then  $v$ is informative for determining $u$'s label only if $v$ is Sybil, because any users can follow a benign user. 

\myparatight{Iteratively solving the posterior reputation scores} The posterior reputation scores in Equation~\ref{unified_linsys} are often solved iteratively. Specifically, the posterior reputation scores are initialized to the prior reputation scores. Then, the posterior reputation scores are iteratively updated until convergence as follows:
{
\begin{align}
\label{unified_linsys_update}
{\mathbf{p}}^{(t+1)} = f(\mathbf{q}, \mathbf{W}, \mathbf{p}^{(t)}), 
\end{align}
}%
where $\mathbf{p}^{(t)}$ is the posterior reputation scores in the $t$th iteration. Note that the edge weight matrix $\mathbf{W}$ is fixed during the iterative process. Finally, in LBP-based methods, 
a node is predicted to be positive if  the node's posterior reputation score is positive, otherwise the node is predicted to be negative.

\subsection{Learning Edge Weights as An Optimization Problem}
\label{optimizationproblem}
Our key intuition is that the edge weights and the posterior reputation scores produced by a collective classification method should satisfy the following two goals.

\begin{itemize}
\item {\bf Goal 1.} The posterior reputation scores of the labeled positive nodes and the labeled negative nodes in the training dataset should be large and small, respectively. 

\item {\bf Goal 2.} The edge weights and the posterior reputation scores should be \emph{consistent}. In particular, we use the posterior reputation scores to predict node labels; 
 an edge is predicted to be homogeneous if the two corresponding nodes are predicted to have the same label, otherwise the edge is predicted to be heterogeneous. 
 Consistency means that 
 an edge that is predicted to be homogeneous  should have a positive weight, while an edge that is predicted to be heterogeneous should have a negative weight. 
\end{itemize}

\myparatight{Quantifying Goal 1}
Given the training dataset $L$, we quantify the Goal 1 as finding an edge weight matrix ${\mathbf{W}}$ such that the difference between the posterior reputation scores of the labeled nodes and their labels is minimized. Formally, we aim to find an edge weight matrix ${\mathbf{W}}$ that \emph{minimizes} the following function: 
{
\begin{align} 
\label{loss}
 \textrm{L}({\mathbf{W}}) = \frac{1}{2} \sum_{l \in L} ({p}_l - {y}_l)^2,
\end{align}
}%
where ${p}_l$ is the posterior reputation score of node $l$ and ${y}_l$ is the label of node $l$ in the training dataset, where ${y}_l=1$ if $u$ is positive and ${y}_l=-1$ otherwise. In machine learning, $\textrm{L}({\mathbf{W}})$ is known as a \emph{loss function} over the training dataset.

\myparatight{Quantifying Goal 2} An edge weight $w_{uv}$ is consistent with the posterior reputation scores of $u$ and $v$ when 1)  $u$ and $v$ are predicted to have the same label (i.e.,  $p_u p_v>0$) and $w_{uv}$ is positive, or 2)   $u$ and $v$ are predicted to have different labels (i.e.,  $p_u   p_v < 0$) and $w_{uv}$ is negative. Therefore, to capture Goal 2, we aim to learn a weight matrix $\mathbf{W}$ that \emph{maximizes} the following function:
{
\begin{align}
\label{regularization}
\textrm{C}({\mathbf{W}}) = \sum_{(u,v)\in E} {p}_u {p}_v {w}_{uv},
\end{align}
}%
where $\textrm{C}({\mathbf{W}})$ measures the consistency of a given weight matrix.

\myparatight{Optimization problem} Combining the two goals, we aim to learn a weight matrix ${\mathbf{W}}$ via solving the following optimization problem:
{
\begin{align} 
\label{objfunc}
\min_{{\mathbf{W}}} \ \mathcal{L}({\mathbf{W}}) = \textrm{L}({\mathbf{W}}) - \lambda \textrm{C}({\mathbf{W}}),
\end{align}
}%
where $\lambda>0$ is a hyperparameter to balance the two goals. 

\myparatight{Challenge for solving the optimization problem} It is computationally challenging to solve the optimization problem in Equation~\ref{objfunc}. The reason is that every node's posterior reputation score depends on every edge weight. For instance, we can use the gradient descent method to solve the optimization problem. Specifically, we can iteratively update  the edge weight $w_{uv}$ as follows:
\begin{align}
\label{gradientdescent}
w_{uv} \leftarrow w_{uv} - \gamma \frac{\partial \mathcal{L}({\mathbf{W}})}{\partial {w_{uv}}}, 
\end{align}
where $\gamma$ is called \emph{learning rate}. However, the derivative $\frac{\partial \mathcal{L}({\mathbf{W}})}{\partial {w_{uv}}}$ depends on $\frac{\partial {\mathbf{p}}}{\partial {w_{uv}}}$ (derivative of every node's posterior reputation score with respect to the edge weight), due to the consistency term $\textrm{C}({\mathbf{W}})$. Since  $\mathbf{p}$ is a solution to Equation~\ref{unified_linsys}, $\frac{\partial {\mathbf{p}}}{\partial {w_{uv}}}$ is a solution to the following system of equations: 
\begin{align}
\label{derivative}
\frac{\partial {\mathbf{p}}}{\partial {w_{uv}}}=\frac{\partial f(\mathbf{q}, \mathbf{W}, \mathbf{p})}{\partial w_{uv}}. 
\end{align}

Therefore, in each iteration of gradient descent at updating the weight matrix, we need to iteratively solve Equation~\ref{derivative} with $|V|$ variables for \emph{each} edge, which is computationally infeasible for large graphs.

\begin{figure}[t]
	\centering
	\includegraphics[width=0.35\textwidth]{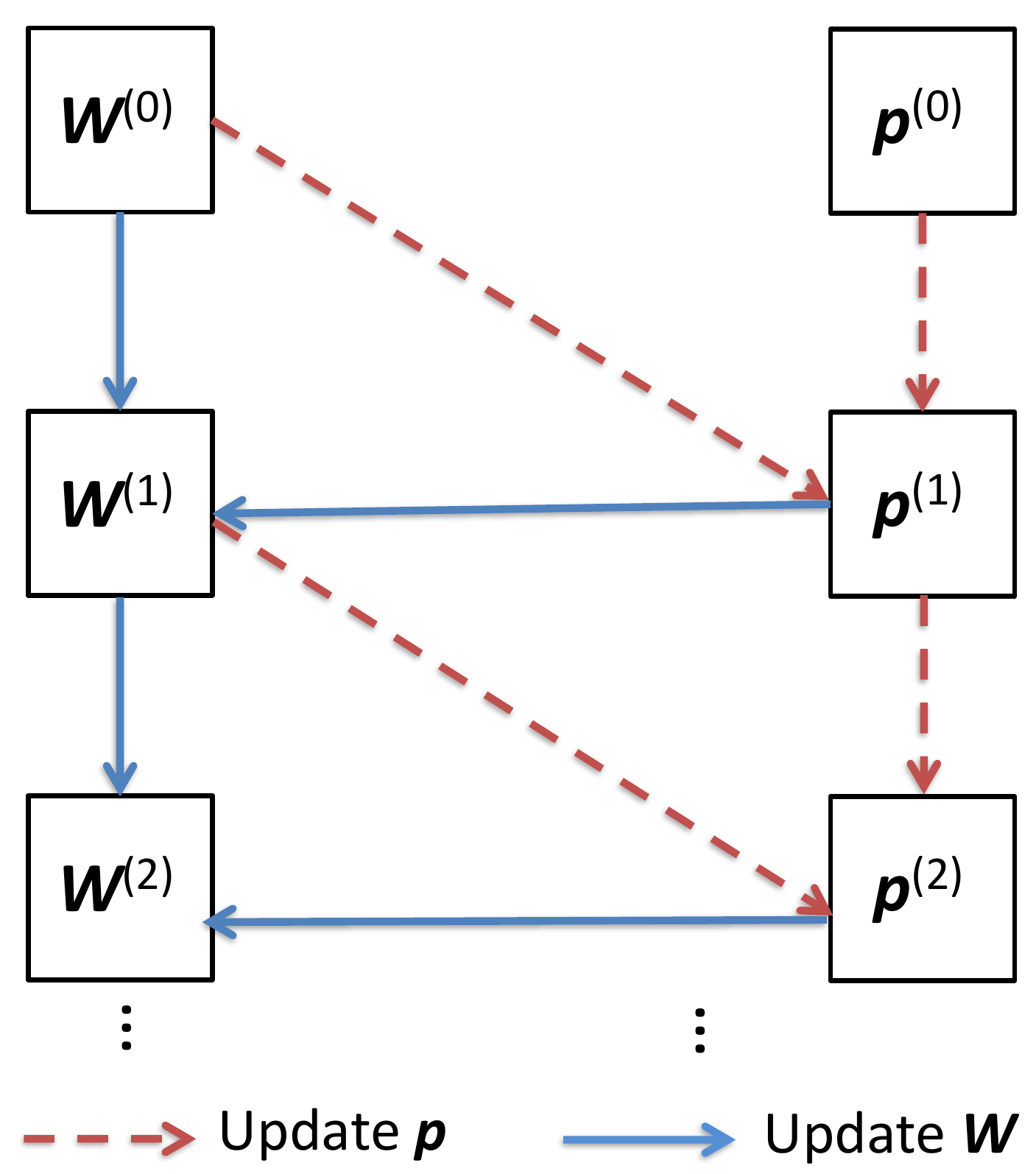}
	\caption{Jointly learning edge weights and updating posterior reputation scores.}
	\label{weightLearn}
\end{figure}

\subsection{Joint Weight Learning and Propagation}
The reason for the computational challenge is that we quantify the two goals using the final posterior reputation scores. 
We observe that the final posterior reputation scores are iteratively computed using Equation~\ref{unified_linsys_update}. Therefore, we propose to quantify the two goals using the current posterior reputation scores during the iterative process. Specifically, given the posterior reputation scores in the $t$th iteration, we aim to learn a weight matrix such that 1) the posterior reputation scores in the $(t+1)$th iteration of the labeled positive nodes and the labeled negative nodes in the training dataset are large and small, respectively (Goal 1); and 2) the edge weights and the posterior reputation scores in the $t$th iteration are consistent (Goal 2). Then, we compute the posterior reputation scores in the $(t+1)$th iteration using the learnt weight matrix. 
In other words, we alternately learn edge weights and propagate reputation scores. Figure~\ref{weightLearn} illustrates our framework.

\myparatight{Propagating posterior reputation scores} Given the weight matrix ${\mathbf{W}}^{(t-1)}$ and the posterior reputation scores ${\mathbf{p}}^{(t-1)}$  in the $(t-1)$th iteration, we compute the posterior reputation score ${\mathbf{p}}^{(t)}$ in the $t$th iteration as follows:
{
\begin{align} 
\label{propgatestep}
&{\mathbf{p}}^{(t)} = f({\mathbf{q}}, {\mathbf{W}}^{(t-1)}, {\mathbf{p}}^{(t-1)}). 
\end{align}
}

\myparatight{Learning weight matrix} Moreover, given the weight matrix ${\mathbf{W}}^{(t-1)}$ in the $(t-1)$th iteration and the posterior reputation score ${\mathbf{p}}^{(t)}$ in the $t$th iteration, we learn the weight matrix ${\mathbf{W}}^{(t)}$ in the $t$th iteration as a solution to the following optimization problem: 
{
\begin{align} 
\label{weightlearning}
&\min_{\mathbf{W}^{(t)}} \mathcal{L}({\mathbf{W}^{(t)}}) = \frac{1}{2} \sum_{l \in L} ({p}_l^{(t+1)} - {y}_l)^2 - \lambda \sum_{(u,v)\in E} {p}_u^{(t)} {p}_v^{(t)} {w}_{uv}^{(t)}, 
\end{align}
}%
where the first term in the objective function $\mathcal{L}({\mathbf{W}^{(t)}})$ quantifies the Goal 1, the second term quantifies the Goal 2, and ${\mathbf{p}}^{(t+1)} = f({\mathbf{q}}, \mathbf{W}^{(t)}, {\mathbf{p}}^{(t)})$. From a machine learning perspective, the first term is known as a \emph{loss function} over the training dataset, while the second term is a regularization term, which we call \emph{consistency regularization}. 
We use gradient descent to solve the optimization problem. Specifically, we initialize ${\mathbf{W}^{(t)}}$ to be the weight matrix ${\mathbf{W}^{(t-1)}}$, and then we iteratively update the edge weight $w_{uv}^{(t)}$ for each edge $(u,v)$ using Equation~\ref{gradientdescent}: $w_{uv}^{(t)} \leftarrow w_{uv}^{(t)} - \gamma \frac{\partial \mathcal{L}({\mathbf{W}^{(t)}})}{\partial {w_{uv}^{(t)}}}$. However, different from the objective function $\mathcal{L}({\mathbf{W}})$ in Equation~\ref{objfunc}, the derivative $\frac{\partial \mathcal{L}({\mathbf{W}^{(t)}})}{\partial {w_{uv}^{(t)}}}$ can be efficiently computed because ${\mathbf{p}}^{(t)}$ is given. In particular, we have:
\begin{align}
\frac{\partial \mathcal{L}({\mathbf{W}^{(t)}})}{\partial {w_{uv}^{(t)}}}=\sum_{l \in L} ({p}_l^{(t+1)} - {y}_l)\frac{\partial {p}_l^{(t+1)}}{\partial {w_{uv}^{(t)}}} - \lambda  {p}_u^{(t)} {p}_v^{(t)}.
\end{align}

For LBP-based methods on undirected graphs, the function $f$ is described in Equation~\ref{sybilscar}. Therefore, we have:
\begin{align}
\text{\bf LBP for undirected graphs:} \nonumber \\
\frac{\partial {p}_l^{(t+1)}}{\partial {w_{uv}^{(t)}}} = 
\begin{cases}
{p}_v^{(t)} &\text{ if } u= l \\
{p}_u^{(t)} &\text{ if } v= l \\
0 &\text{ otherwise} 
\end{cases}
\end{align}

For LBP on directed graphs, the function $f$ is described in Equation~\ref{gang}. Therefore, we have:
{\small
\begin{align}
&\text{\bf LBP for directed graphs:} \nonumber \\
&\frac{\partial {p}_l^{(t+1)}}{\partial {w_{uv}^{(t)}}} = 
\begin{cases}
A_{b,uv} {p}_v^{(t)} + A_{i,uv}  \mathcal{I}({p}_v^{(t)}) + A_{o,uv}  \mathcal{J}({p}_v^{(t)})  &\text{ if } u= l \\
0 &\text{ otherwise}
\end{cases}
\end{align}
}%
Note that, in gradient descent, we can repeat multiple iterations to compute the weight matrix $\mathbf{W}^{(t)}$. 
\alan{However, we find that our methods already work well using only one iteration. Moreover, one iteration makes our methods more efficient. 
Therefore, we will apply one iteration to compute $\mathbf{W}^{(t)}$.}

We alternately propagate posterior reputation scores and learn edge weights until the posterior reputation scores in two consecutive alternations are small enough (e.g., $\| {\mathbf{p}}^{(t+1)} - {\mathbf{p}}^{(t)} \|_1 / \| {\mathbf{p}}^{(t+1)} \|_1 < 10^{-3}$) or we have reached the allowed maximum number of alternations (e.g., 15 in our experiments). 

\myparatight{Computational complexity} 
In both propagating posterior reputation scores and learning edge weights, our framework traverses all edges. The time complexity of our framework is $O( |E| \cdot T)$, where $T$ is the number of alternations. State-of-the-art LBP-based methods~\cite{pandit2007netprobe,chau2011polonium,sybilbelief,wang2017sybilscar,jia2017attriinfer,wang2017gang} have the same asymptotic time complexity. As we will demonstrate in our experiments, our framework will be 2-3 times slower than state-of-the-art LBP-based methods in practice. This is because our framework learns edge weights in each iteration. However, this time overhead is tolerable in practice, especially the targeted security and privacy applications are not time-critical. For instance, on a Twitter dataset with 1.5 billion edges, our method finishes within 3 hours on a server with 512GB memory and 32 cores. Moreover, our method can be easily parallelized and should be scalable to graphs with tens of billions of edges on a modern data center.

\section{Evaluation}
\label{exp}

We evaluate our framework on two security applications (Sybil detection, fake review detection) and one privacy application (attribute inference attacks in social networks). 
We compare our framework with state-of-the-art RW-based methods and LBP-based methods in terms of AUC, scalability, and robustness to label noise.

\subsection{Experimental Setup}

\subsubsection{Dataset Description}
We use a Twitter dataset and a Sina Weibo dataset for Sybil detection, a Yelp dataset for fake review detection, and a Google+ dataset for attribute inference attacks. Table~\ref{dataset_stat} summarizes the basic dataset statistics.

\myparatight{Twitter} We obtained a Twitter dataset with real Sybils from Wang et al.~\cite{wang2017sybilscar}. Specifically, the directed Twitter follower-followee graph was collected by  Kwak et al.~\cite{kwak2010twitter}. A directed edge $(u,v)$ means that $u$ follows $v$. This graph contains around 42M users and 1.5B edges. 
A user that was suspended by Twitter is treated as a Sybil (positive), while an active user is treated as benign (negative). In total, 205,355 nodes are Sybils, 36,156,909 nodes are benign, and the remaining nodes are unlabeled.

\myparatight{Sina Weibo} We obtained a Sina Weibo dataset with around 3.5M users and 653M directed edges from Fu et al.~\cite{robustspammer}. 
Like Twitter, a directed edge $(u,v)$ means that $u$ follows $v$. 
Fu et al. also manually labeled 2000 users sampled uniformly at random. 
Among them, 482 were Sybil users, 1498 were benign users, and 20 were unknown. 

\myparatight{Yelp} We obtained a Yelp dataset from~\cite{rayana2015collective}, which contains Yelp reviews for restaurants located in New York City. 
The dataset has 160,225 users, 923 products, and 359,052 reviews. By constructing an undirected User-Review-Product graph, we have 520,230 nodes and 718,144 edges. 
A review is treated as fake if the review is filtered or not recommended by Yelp, otherwise the review is treated as genuine.
Around 10\% of the reviews are fake in the dataset. In the User-Review-Product graph, we only concern the labels of the review nodes, where a fake review is treated as a positive node and a genuine review is treated as a negative node. 

\myparatight{Google+} We obtained an undirected Google+ social network with user attributes from~\cite{Gong12-imc,jia2017attriinfer}. The dataset has around 5.7M users and 31M edges. The user attribute is  \emph{cities lived} and the dataset has 50 most popular cities. 3.25\% of users disclosed at least one of these cities as their cities lived. We treat each city as a binary attribute and perform attribute inference for each binary attribute separately.  Specifically, given a city, a user is treated as positive if the user lives/lived in the city, while treated as negative if the user does not.

\begin{table}[!t]\renewcommand{\arraystretch}{1.8}
\centering
\caption{Dataset statistics.}
\centering
\begin{tabular}{|c|c|c|c|} \hline 
{\small Dataset} & {\small \#Nodes} & {\small \#Edges} &{\small Ave. degree}\\ \hline
{\small Twitter} &  {\small 41,652,230} & {\small 1,468,364,884} & {\small 71}  \\ \hline
{\small Sina Weibo} &  {\small 3,538,487} & {\small 652,889,971} & {\small 369} \\ \hline
{\small Yelp} & {\small 520,230} & {\small 718,144} & {\small 3}\\ \hline
{\small Google+} & {\small 5,735,175} & {\small 30,644,909} & {\small 11}\\ \hline
\end{tabular}
\label{dataset_stat}
\end{table}

\subsubsection{Training and testing}
On  Twitter, we sample 3,000 positive nodes and 3,000 negative nodes uniformly at random as the training set; and we treat the remaining positive and negative nodes as the testing set.
On Sina Weibo, we split the 1980 labeled users into two halves; one is treated as the training set and the other as the testing set.
On  Yelp, we randomly sample 1,000 fake reviews and  1,000 genuine reviews as the training set; and we treat the remaining reviews as the testing set. 
On  Google+, we select 75\% of the users who have at least one city as {training set} and treat the remaining users who have at least one city as {testing set}. 
Recall that we have 50 cities in the Google+ dataset, where each city is treated as a binary attribute independently. 
The AUC of one method for the Google+ dataset is averaged over the 50 cities.

{\small
\begin{table}[!t]\renewcommand{\arraystretch}{2}
\centering
\caption{AUCs of undirected methods.}
\label{AUC_undirect}
\addtolength{\tabcolsep}{-2pt}
\begin{tabular}{|c|c|c|c|c|c|}
\hline
\multicolumn{2}{|c|}{\small \textbf{Methods}} & {\small \textbf{Twitter}}  &  {\small \textbf{Sina Weibo}} &   {\small \textbf{Yelp}} &  {\small \textbf{Google+}} \\ \hline
\multirow{4}{*}{\bf RW} & {\small RW-N-U} & {\small 0.57}  & {\small 0.61} &   {\small 0.55} &   {\small 0.59}  \\ \cline{2-6}
{\bf } & {\small RW-P-U} & {\small 0.58}  & {\small 0.61} &   {\small 0.57} &   {\small 0.58}  \\ \cline{2-6}
{\bf } &{\small RW-LFW-U} & {\small 0.53}  & {\small 0.54} &   {\small 0.48}  &   {\small 0.57} \\ \cline{2-6} 
{\bf } & {\small RW-B-U} & {\small 0.63} &   {\small 0.68} &   {\small 0.58} &   {\small 0.63}  \\ \hline  \hline
\multirow{2}{*}{\bf LBP} & {\small LBP-U} &  {\small 0.64}  &  {\small 0.68}  &   {\small 0.58} &   {\small 0.66} \\  \cline{2-6}
{\bf } & {\small LBP-FLW-U} &  {\small 0.62}  &  {\small 0.66} &   {\small 0.58} &   {\small 0.66}  \\ \hline  \hline
\multirow{4}{*}{\bf Ours} {\bf } & {\small LBP-JWP-w/o-U} &  {\small {0.69}}  &  {\small {0.74}}  &   {\small 0.60} &   {\small {0.69}}  \\ \cline{2-6}
{\bf } & {\small LBP-JWP-L1-U} &  {\small {0.65}}  &  {\small {0.70}}  &   {\small {0.59}} &   {\small {0.66}}  \\ \cline{2-6}
{\bf } & {\small LBP-JWP-L2-U} &  {\small {0.68}}  &  {\small {0.72}}  &   {\small {0.60}}  &   {\small {0.68}} \\ \cline{2-6}
{\bf } & {\small LBP-JWP-U} &  {\small \textbf{0.73}}  &  {\small \textbf{0.77}}  &   {\small \textbf{0.62}} &   {\small \textbf{0.72}}  \\ \hline
\end{tabular} \\
\end{table}
}%

\subsubsection{Compared methods} 
We compare our framework with state-of-the-art RW-based methods and LBP-based methods on undirected graphs and directed graphs. 

\myparatight{RW-based methods for undirected graphs (RW-N-U, RW-P-U, RW-B-U, and RW-FLW-U)} We consider the RW-based methods RW-N~\cite{sybilrank}, RW-P~\cite{wu2006propagating}, RW-B~\cite{jia2017random}, and RW-FLW~\cite{integro}. Please refer to Section~\ref{relatedwork} for details about these methods. Note that the method developed by Wu et al.~\cite{wu2006propagating} is originally for directed graphs. We apply it to undirected graphs via ignoring the edge directions. RW-FLW requires a classifier to predict the probability that a node is a victim. We consider the classifier is optimal, i.e., it correctly predicts all victims, which  gives advantages to RW-FLW. More specifically, we set such probability of each victim to be 0.9 and such probability of each non-victim to be 0.1. 
 We add a suffix ``-U'' to each method to indicate that they are for undirected graphs. 

We note that Twitter and Sina Weibo are directed graphs. When applying these methods on the two datasets, we transform them to undirected graphs. Specifically, we keep an edge between two nodes only if they are connected by directed edges in both directions. Moreover, these methods only work for connected graphs, so we extract the largest connected component from the transformed undirected graph.

\myparatight{RW-based methods for directed graphs (RW-N-D and RW-P-D)} We use TrustRank~\cite{gyongyi2004combating} as RW-N-D and DistrustRank~\cite{wu2006propagating} as RW-P-D. We add a suffix ``-D'' to indicate that these methods are for directed graphs.

\myparatight{LBP-based methods for undirected graphs (LBP-U, LBP-FLW-U)} We consider the LBP method with the optimized LBP~\cite{jia2017attriinfer} and LBP-FLW~\cite{gao2018sybilfuse} that learns edge weights. Like RW-based methods for undirected graphs, we transform the directed Twitter and Sina Weibo graphs into undirected ones. In LBP-FLW, we extract structure features for nodes and learn an SVM classifier to predict the prior reputation score for each unlabeled node. Then, following the setting in~\cite{gao2018sybilfuse}, we assign a weight $0.4$ to an edge if the two corresponding nodes are predicted to have the same label and a weight $-0.4$ otherwise. For the undirected Twitter and Sina Weibo graphs, the features are extracted from the original directed graph and include local clustering coefficient, incoming edges accepted ratio, and outgoing edges accepted ratio~\cite{gao2018sybilfuse}. For the undirected Yelp and Google+ graphs, only the local clustering coefficient is used.

\myparatight{LBP-based methods for directed graphs (LBP-D)} Only one LBP-based method~\cite{wang2017gang} for directed graphs was developed and we compare with it. 

\myparatight{Our LBP-based methods (LBP-JWP-w/o, LBP-JWP-L1, LBP-JWP-L2, and LBP-JWP)} LBP-JWP-w/o is the variant of our LBP-based method that does not use the consistency term in the optimization problem in Equation~\ref{weightlearning} when learning edge weights. From the machine learning perspective, the first term in the objective function of the optimization problem in Equation~\ref{weightlearning} is a \emph{loss function} over the training dataset, and the second term  is a regularization term about the edge weights, which we call \emph{consistency regularization}. In machine learning, $L_1$ and $L_2$ regularizations are widely used, which are $-\sum_{(u,v)\in E} |w_{uv}|_1$ and $-\sum_{(u,v)\in E} w_{uv}^2$, respectively. LBP-JWP-L1 and LBP-JWP-L2 respectively are the variants that use the conventional $L_1$ and $L_2$ regularizations instead of our consistency regularization when learning edge weights. LBP-JWP is our method with the consistency regularization. We add the suffix ``-U'' and ``-D'' to indicate the versions that are used for undirected and directed graphs, respectively.

\begin{table}[!t]\renewcommand{\arraystretch}{2}
\centering
\caption{AUCs of directed methods.}
\label{AUC_direct}
\begin{tabular}{|c|c|c|c|}
\hline
 \multicolumn{2}{|c|}{\small \textbf{Methods}} & {\small \textbf{Twitter}}  &  {\small \textbf{Sina Weibo}} \\ \hline
\multirow{2}{*}{\bf RW}& {\small RW-N-D} & {\small 0.60}   &  {\small 0.66} \\ \cline{2-4}
{\bf } & {\small RW-P-D} & {\small 0.63}    & {\small 0.64} \\ \hline  \hline
{\bf LBP} & {\small LBP-D} & {\small {0.72}} & {\small {0.80}} \\ \hline  \hline
\multirow{4}{*}{\bf Ours} & {\small LBP-JWP-w/o-D} & {\small {0.75}}  & {\small {0.82}} \\ \cline{2-4}
{\bf } & {\small LBP-JWP-L1-D} & {\small {0.72}}  & {\small {0.79}} \\ \cline{2-4}
{\bf } & {\small LBP-JWP-L2-D} & {\small {0.73}}  & {\small {0.80}} \\ \cline{2-4}
{\bf } & {\small LBP-JWP-D} & {\small \textbf{0.78}}  & {\small \textbf{0.85}} \\ \hline
\end{tabular} \\
\vspace{-2mm}
\end{table}

\begin{figure*}[!t]
\center
\subfloat[LBP-JWP-U]{\includegraphics[width=0.42\textwidth]{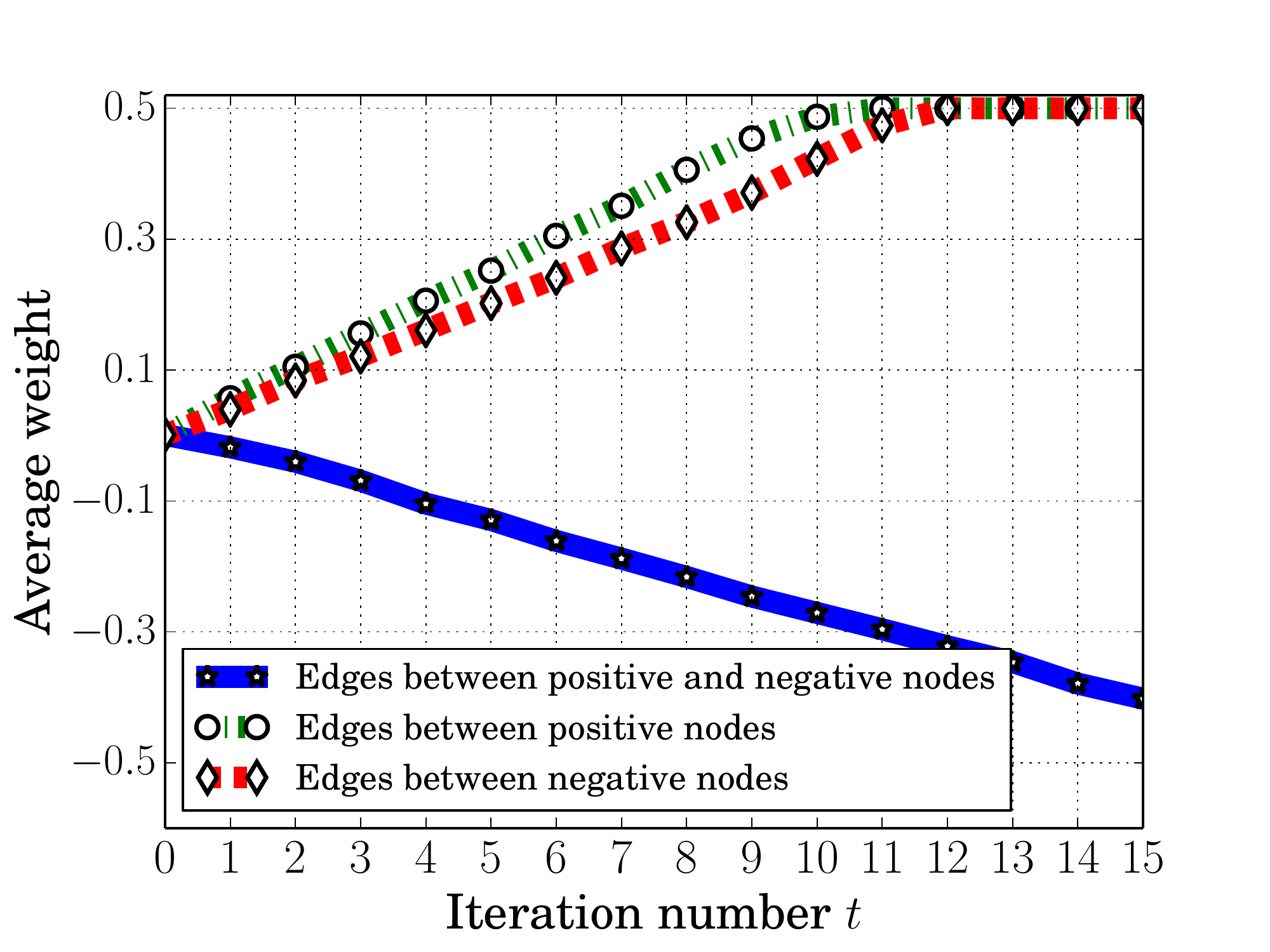}}
\subfloat[LBP-JWP-D]{\includegraphics[width=0.42\textwidth]{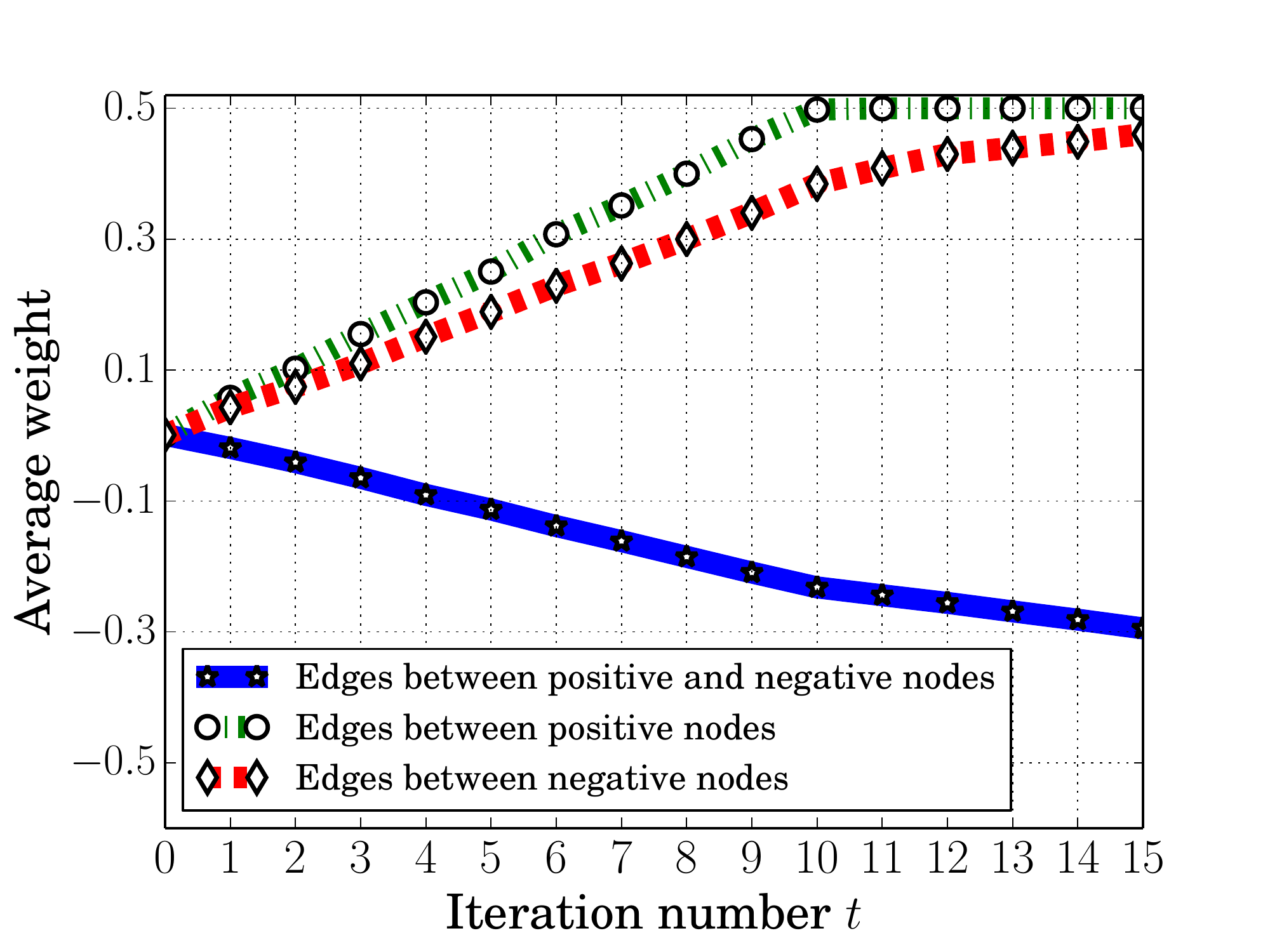}}
\caption{Average weight learnt by (a) LBP-JWP-U and (b) LBP-JWP-D of edges between positive nodes and negative nodes, edges between positive nodes, and edges between negative nodes on Twitter, as the number of iterations increases. We observe that on both methods, the average weights of edges between positive nodes and negative nodes decrease, while the average weights of edges between negative (or positive) nodes increase as $t$ increases.}
\label{homo_stre}
\vspace{-6mm}
\end{figure*}

\subsubsection{Implementation and parameter setting} We implement the RW-based methods and our methods in C++. The authors of the existing LBP-based methods~\cite{wang2017sybilscar,wang2017gang} made their C++ source code publicly available, and we use them to evaluate existing LBP-based methods. 
\alan{We performed all our experiments on a Linux machine with 512GB memory and 32 cores.}
For our methods, 
we assign  prior reputation scores 1, -1, and 0 to labeled positive nodes, labeled negative nodes, and unlabeled nodes, respectively. 
\alan{Following~\cite{wang2017sybilscar} and~\cite{wang2017gang}, we initialize all edge weights to be the inverse of the average node degree in a graph, i.e., $w^{(0)}_{uv} = \frac{1}{\textrm{Ave. degree}}$.
Moreover, we normalize edge weights to be in the range [-0.5,0.5] in each iteration. 
We set the learning rate $\gamma$ in Equation~\ref{gradientdescent} to be 1.0. 
Moreover, we set the regularization parameter $\lambda=0.1$ on the Twitter and Sina Weibo datasets and $\lambda=1.0$ on the Google+ and Yelp datasets considering their different average node degrees. 
Note that we also explore the impact of $\gamma$ and $\lambda$ and show the results in Figure~\ref{impact_hyper}.}
For all other compared methods, we set the parameters according to their authors.

\subsection{Accuracy (AUC)}

Table~\ref{AUC_undirect} and Table~\ref{AUC_direct} respectively show the AUCs of the compared undirected graph and directed graph based methods. 

\myparatight{Our methods outperform existing ones} Our methods LBP-JWP-U and LBP-JWP-D achieve the best AUCs consistently on different datasets. For instance, on undirected graphs, LBP-JWP-U improves upon the best existing method by 0.04 to 0.09 on the four datasets. This means that our strategy of jointly learning edge weights and propagating reputation scores is effective.  
Moreover,  LBP-JWP outperforms LBP-JWP-w/o, LBP-JWP-$L_1$, and LBP-JWP-$L_2$ on both undirected and directed graphs. This means that our consistency regularization does improve AUC and outperforms conventional $L_1$ and $L_2$ regularizations. In fact,  $L_1$ and $L_2$ regularizations decrease AUCs, compared to LBP-JWP-w/o that does not use regularization. 

\alan{Note that all compared methods have relatively low AUCs on Yelp. This is because the Yelp graph is very sparse, i.e., Ave. degree=3, and thus its graph structure has less information that can be exploited by graph-based methods.}

\begin{figure*}[!t]
\center
\subfloat[Impact of $\lambda$]{\includegraphics[width=0.42\textwidth]{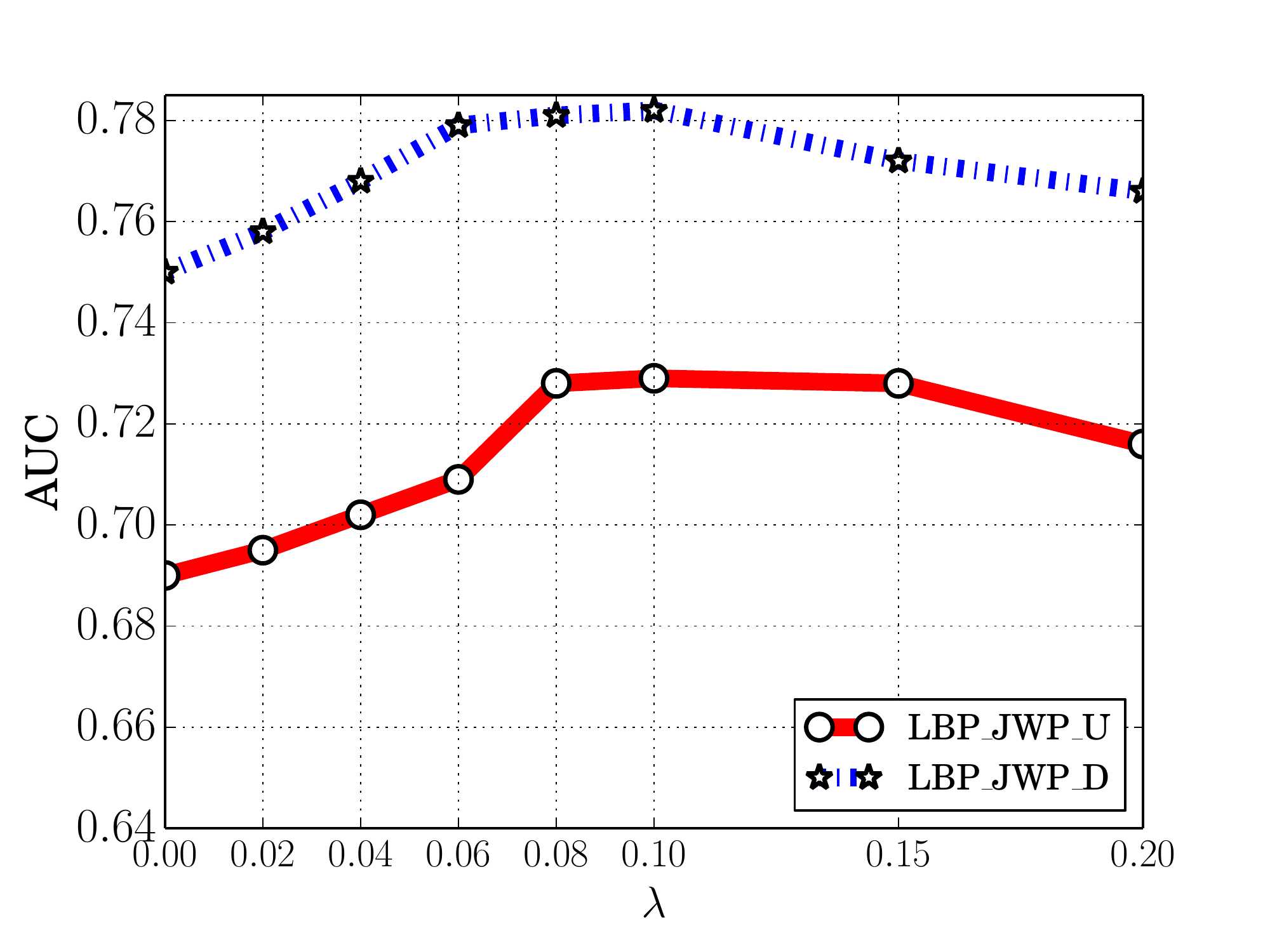} \label{impact_lambda}} 
\subfloat[Impact of $\gamma$]{\includegraphics[width=0.42\textwidth]{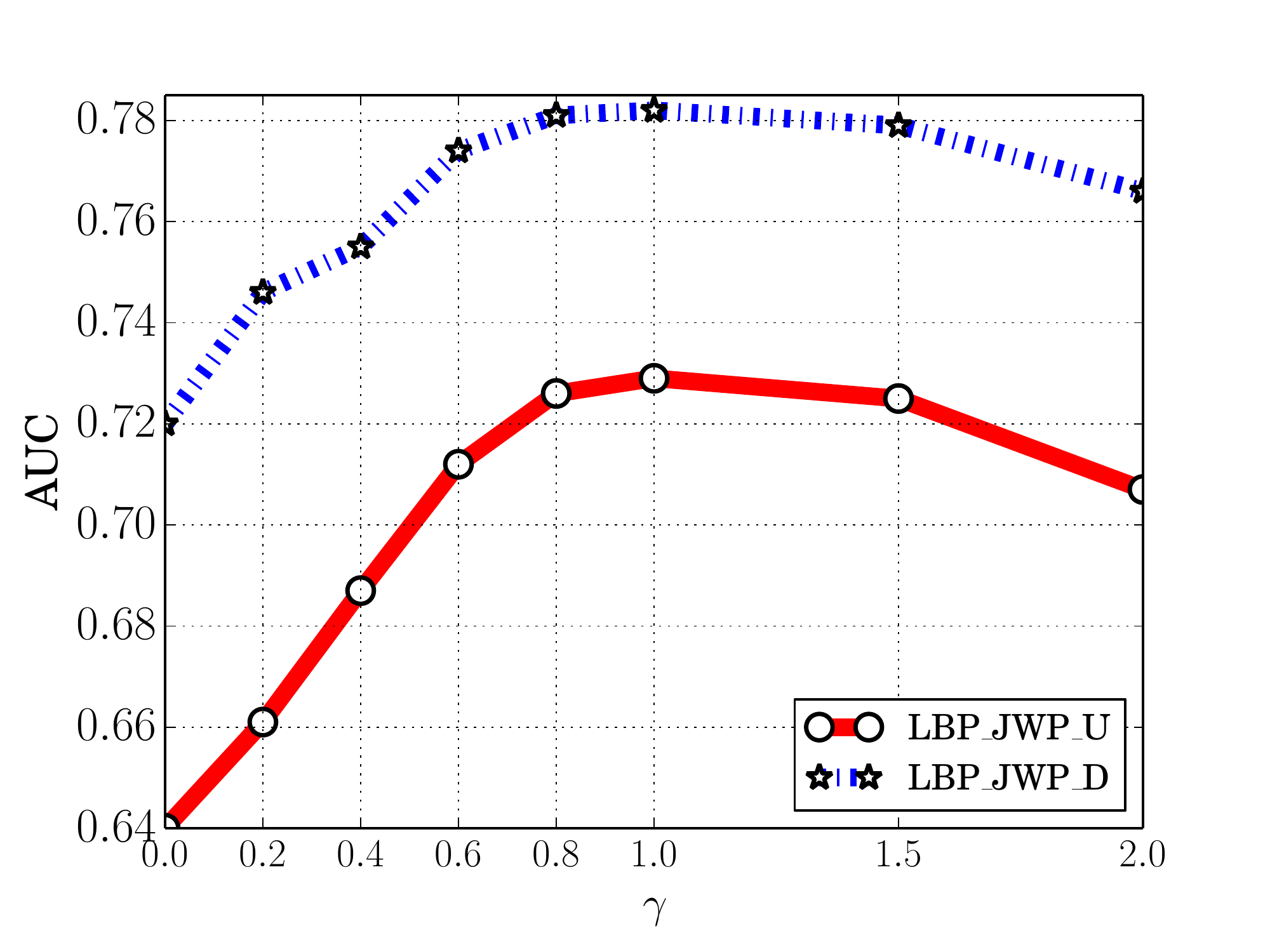} \label{impact_gamma}}
\caption{Impact of (a) the regularization parameter $\lambda$ and (b) the learning rate $\gamma$. We observe that, as $\lambda$ and $\gamma$ increase, AUCs of our methods first increase, then stabilize, and finally decrease.}
\label{impact_hyper}
\vspace{-2mm}
\end{figure*}

\begin{figure}[!t]
\vspace{-5mm}
\center
{\includegraphics[width=0.42\textwidth]{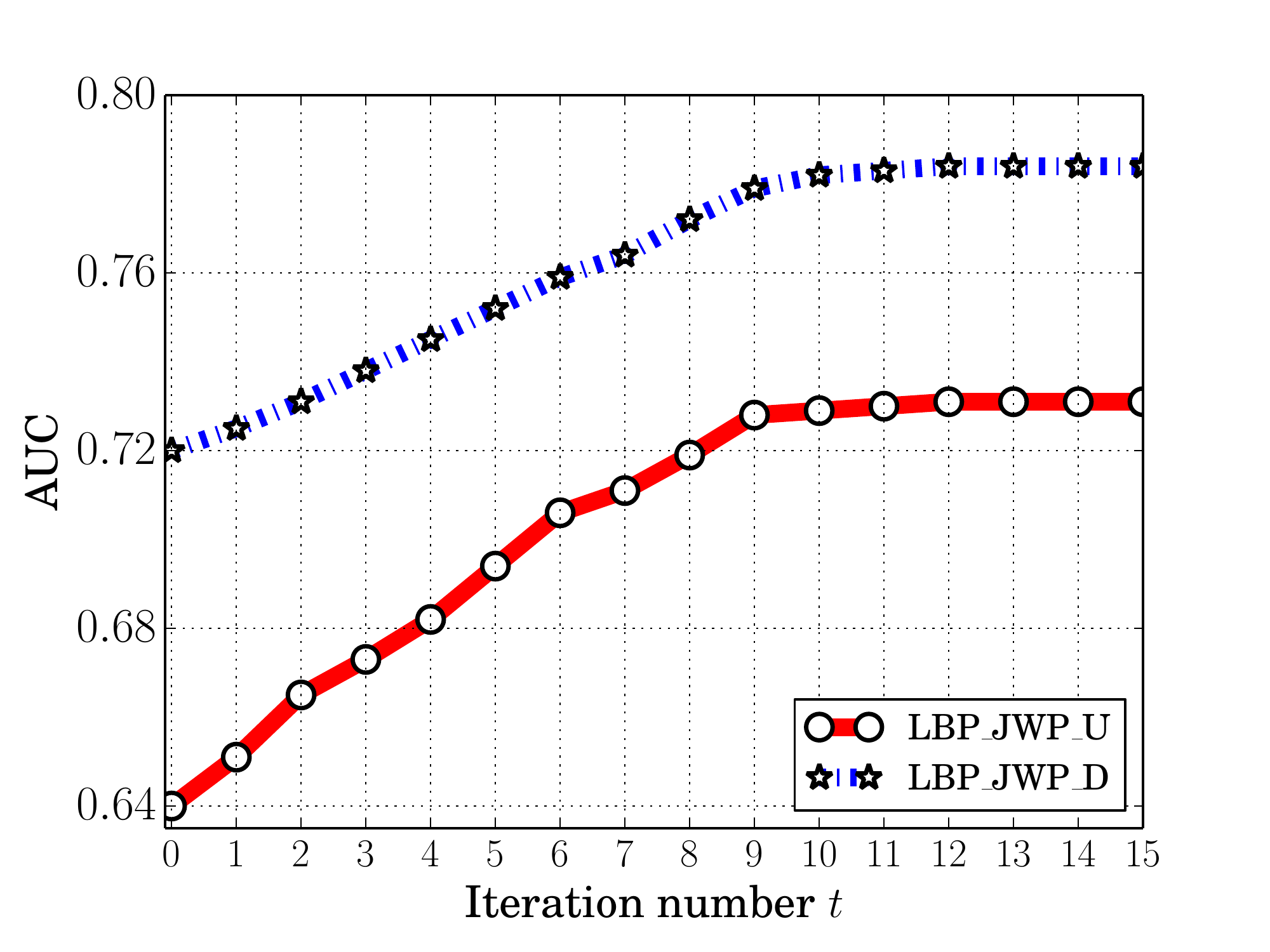}}
\caption{Impact of the number of iterations on the AUCs of our methods. We observe that AUCs first gradually increase as we perform more iterations and then stabilize.}
\label{impact_iternum}
\vspace{-6mm}
\end{figure}

To further understand why our framework outperforms existing methods, we visualize the learnt edge weights. Specifically,  
we classify edges into three categories: 1) edges between positive nodes and negative nodes, 2) edges between positive nodes, and 3) edges between negative nodes. The edges in the first category are heterogeneous, while the edges in the second and third categories are homogeneous. Figure~\ref{homo_stre} shows the average weights of the three categories of edges as a function of the number of iterations $t$ on the Twitter dataset. As we can see, for both LBP-JWP-U and LBP-JWP-D,  the average weights of edges between positive nodes and negative nodes decrease, while the average weights of edges between negative (or positive) nodes increase as the number of iterations increases. This is because we enforce the consistency regularization when learning the edge weights. 

However, for most existing methods, the edge weights are constant, which ignore the differences between heterogeneous and  homogeneous edges. RW-FLW learns edge weights. However, we find that the average weights of all three types of edges are close to -0.3 (after normalizing edge weights to be [-0.5, 0.5]). This is because a large number of nodes are victims. Moreover, for LBP-FLW, the average weights of all three types of edges are close to 0.2. 
This means that LBP-FLW can not  learn small weights for heterogeneous edges. Note that all these methods fix the edge weights and propagate reputation scores using the fixed edge weights.

\myparatight{LBP-based methods outperform RW-based methods} We find that LBP-based methods 
achieve at least the same AUCs as RW-based methods, and LBP-based methods achieve higher AUCs in most cases. 
In particular, for methods on undirected graphs, the best existing LBP-based method achieves the same AUCs as the best existing RW-based method on Sina Weibo and Yelp, while the best  existing LBP-based method achieves slightly higher AUCs on Twitter and Google+. For methods on directed graphs, the existing LBP-based method significantly outperforms the existing RW-based methods. For instance, on Sina Weibo, LBP-D outperforms RW-N-D by 0.14. A possible reason is that LBP-D leverages both labeled positive nodes and labeled negative nodes in the training dataset, while RW-N-D only considers labeled negative nodes.  

\myparatight{Different variants of RW-based methods} We observe that RW-B-U consistently outperforms other RW-based methods for undirected graphs. One reason could be that RW-B-U incorporates both labeled positive nodes and labeled negative nodes in the training dataset, while the other RW-based methods incorporate either of them. Performance of RW-N and RW-P is dataset-dependent, i.e., RW-N outperforms RW-P on some datasets while RW-P outperforms RW-N on other datasets.

\alan{
\myparatight{Impact of the regularization parameter $\lambda$ and the learning rate $\gamma$} Figure~\ref{impact_lambda} shows AUCs of our methods vs. the regularization parameter $\lambda$ on the Twitter dataset (we have similar tendencies on the other datasets and thus choose  Twitter for simplicity).  $\lambda=0$ means that we do not use the consistency
term. 
We observe that AUCs first increase, then stabilize, and finally decrease as $\lambda$ increases.  For instance, AUC of LBP-JWP-U increases until $\lambda$ is around 0.08, then stabilizes until $\lambda$ is around 0.15, and finally decreases after that. 
Figure~\ref{impact_gamma} shows AUCs of our methods vs. the learning rate $\gamma$ on the Twitter dataset.  $\gamma=0$ means that we do not learn edge weights, and our methods reduce to existing methods. Likewise, we observe that AUCs first increase, then stabilize, and finally decrease.
}

\alan{
\myparatight{Impact of the number of iterations $t$} Figure~\ref{impact_iternum} shows the AUCs of our methods on Twitter vs. iteration number $t$. We observe that AUCs first gradually increase as our methods run more iterations and then stabilize after around 10 iterations. 
}

\begin{figure*}[!tbp]
\center
\vspace{-4mm}
\subfloat[]{\includegraphics[width=0.42\textwidth]{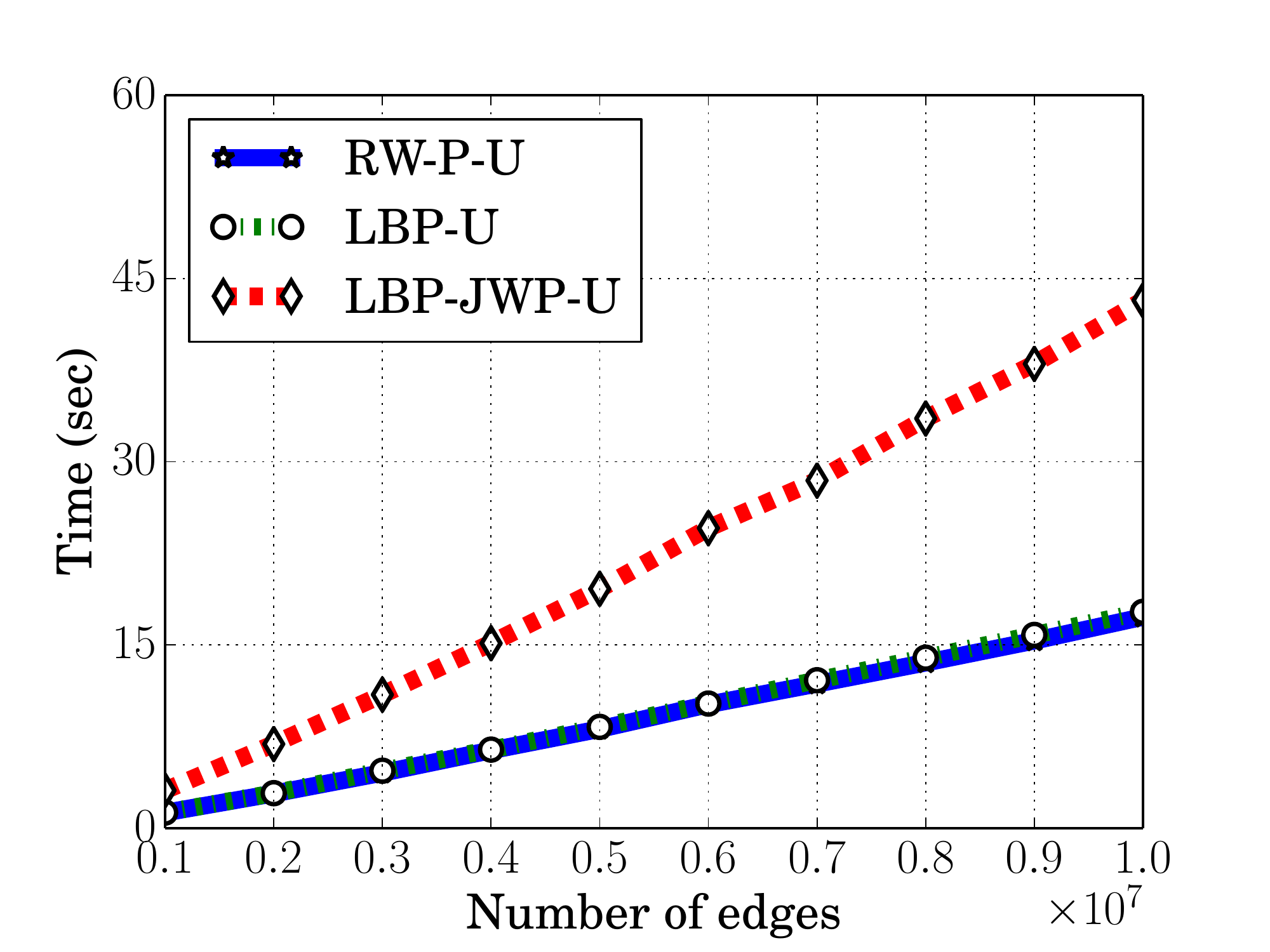} \label{efficiency-u}}
\subfloat[]{\includegraphics[width=0.42\textwidth]{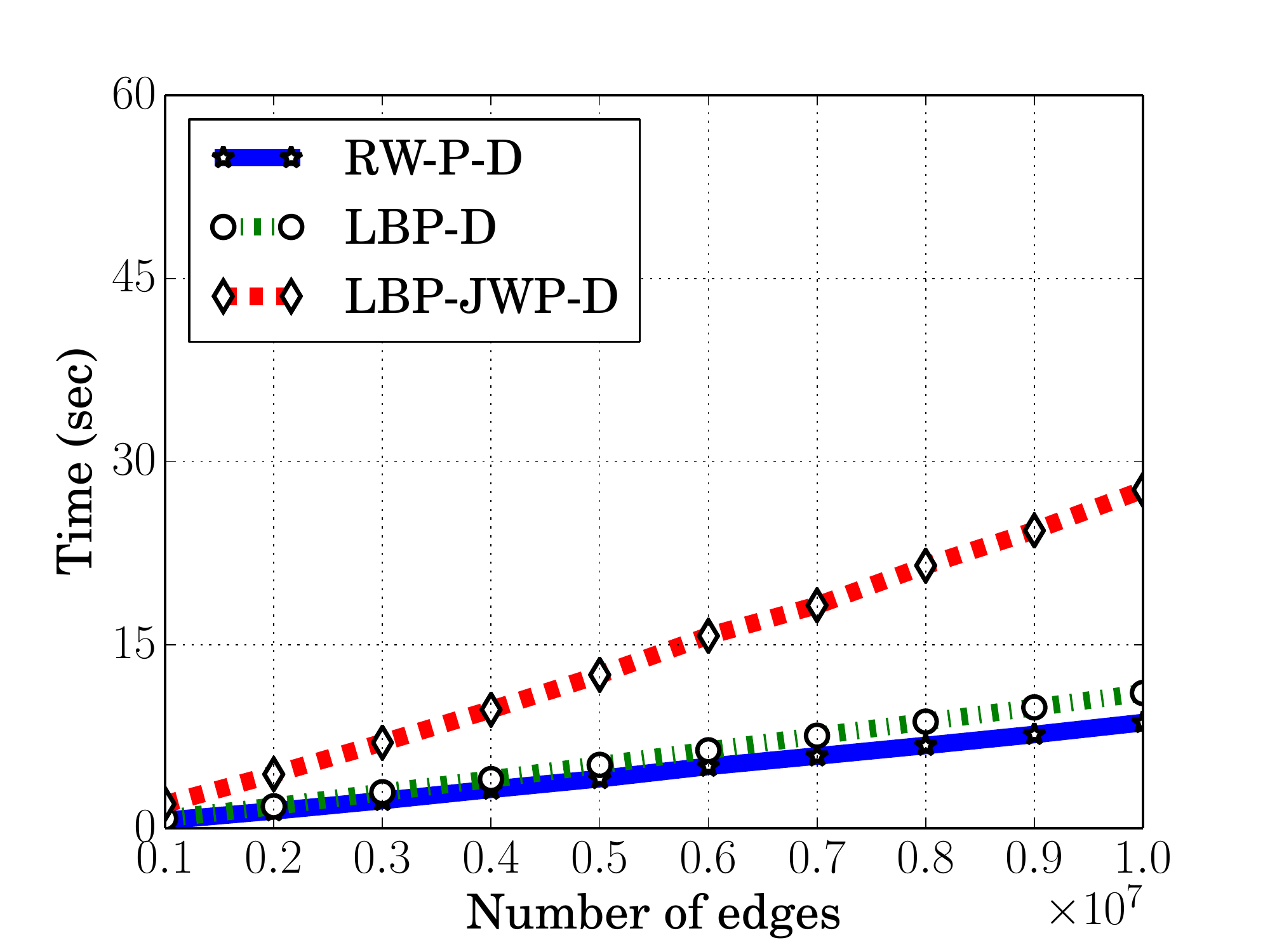} \label{efficiency-d}}
\caption{Running time on synthesized graphs with increasing number of edges. (a) Methods for undirected graphs. (b) Methods for directed graphs.}
\vspace{-6mm}
\end{figure*}

\begin{figure*}[!tbp]
\center
\subfloat[]{\includegraphics[width=0.33\textwidth]{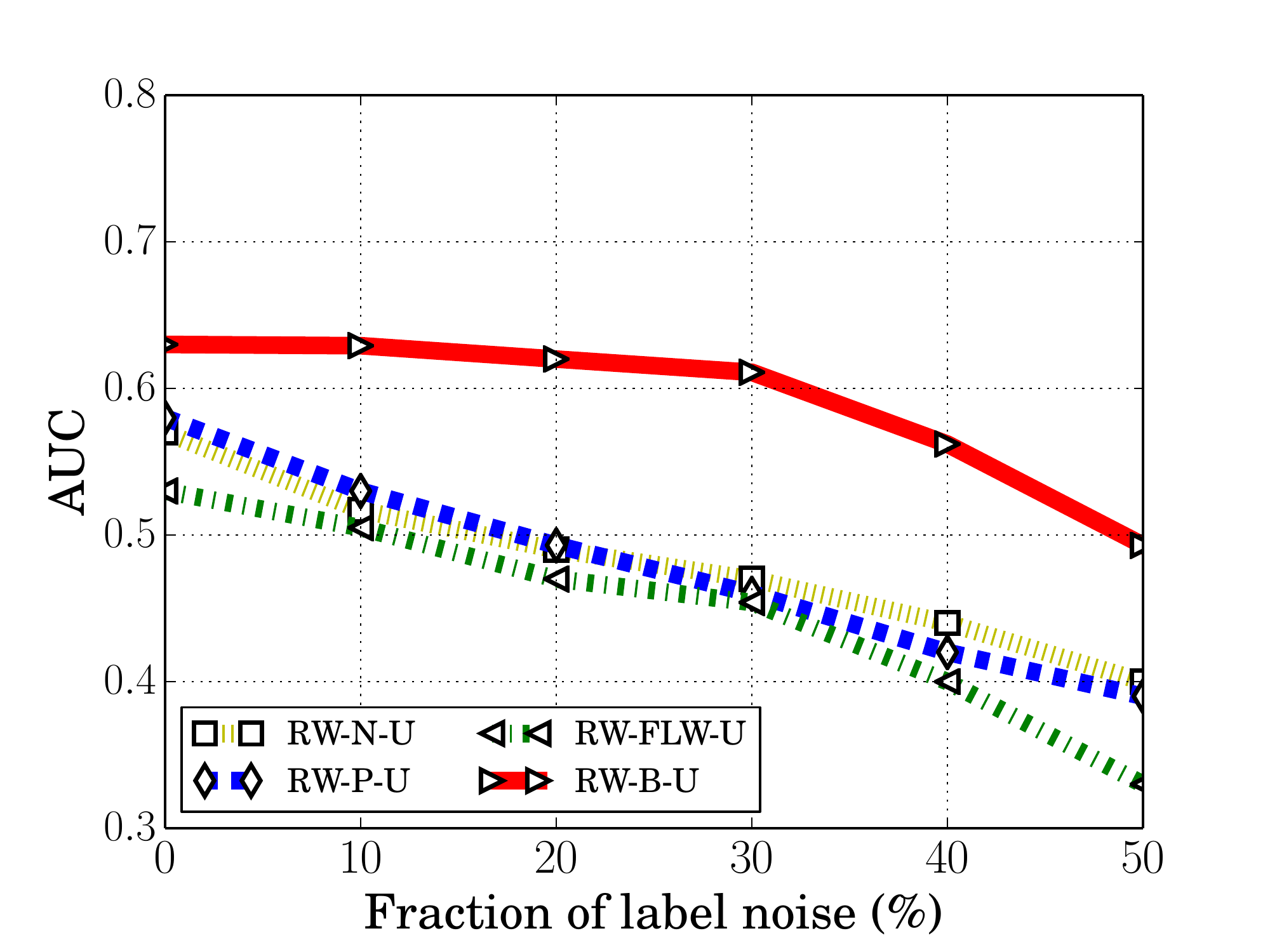} \label{noise-u-rw}}
\subfloat[]{\includegraphics[width=0.33\textwidth]{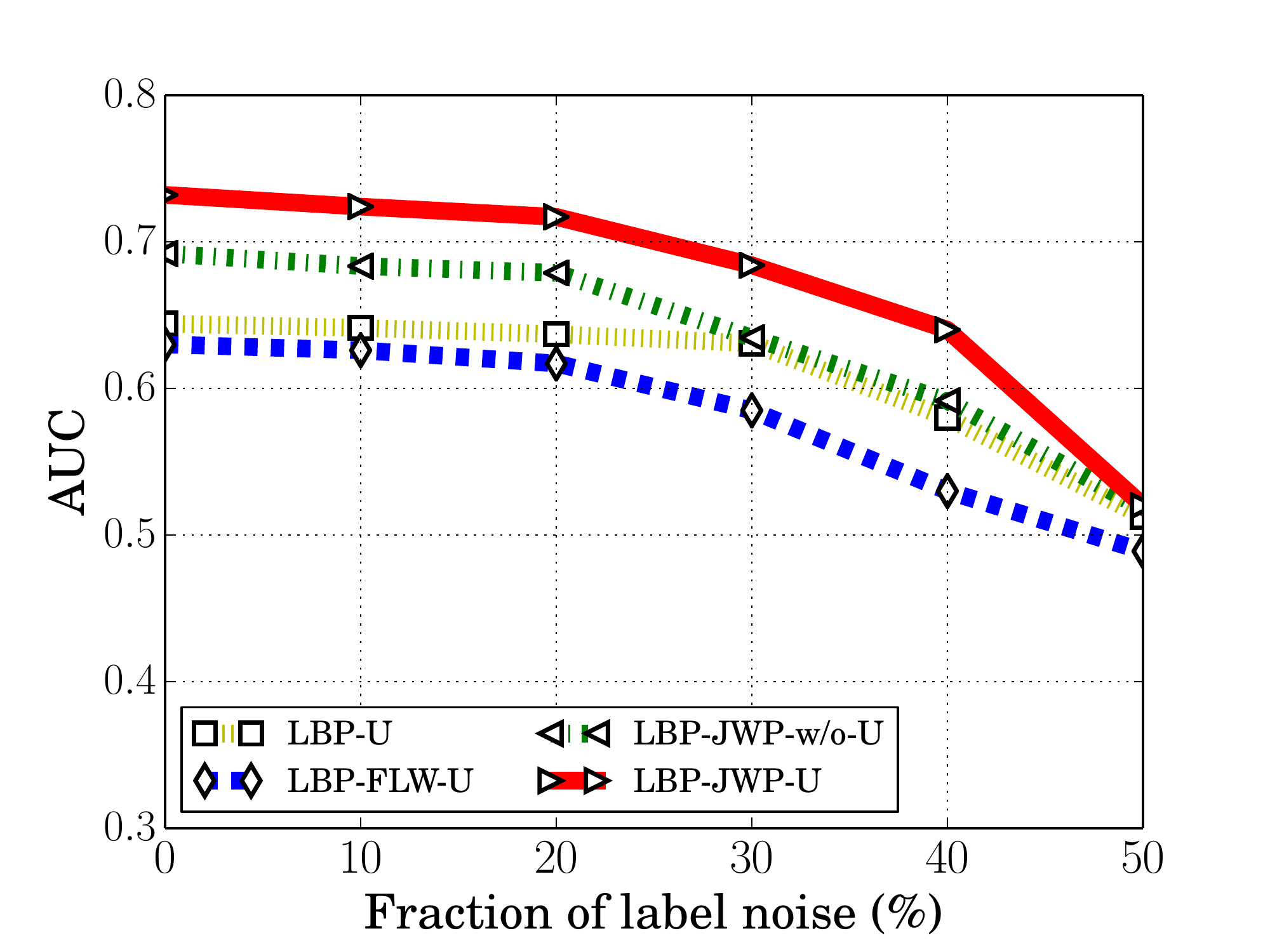} \label{noise-u-lbp}}
\subfloat[]{\includegraphics[width=0.33\textwidth]{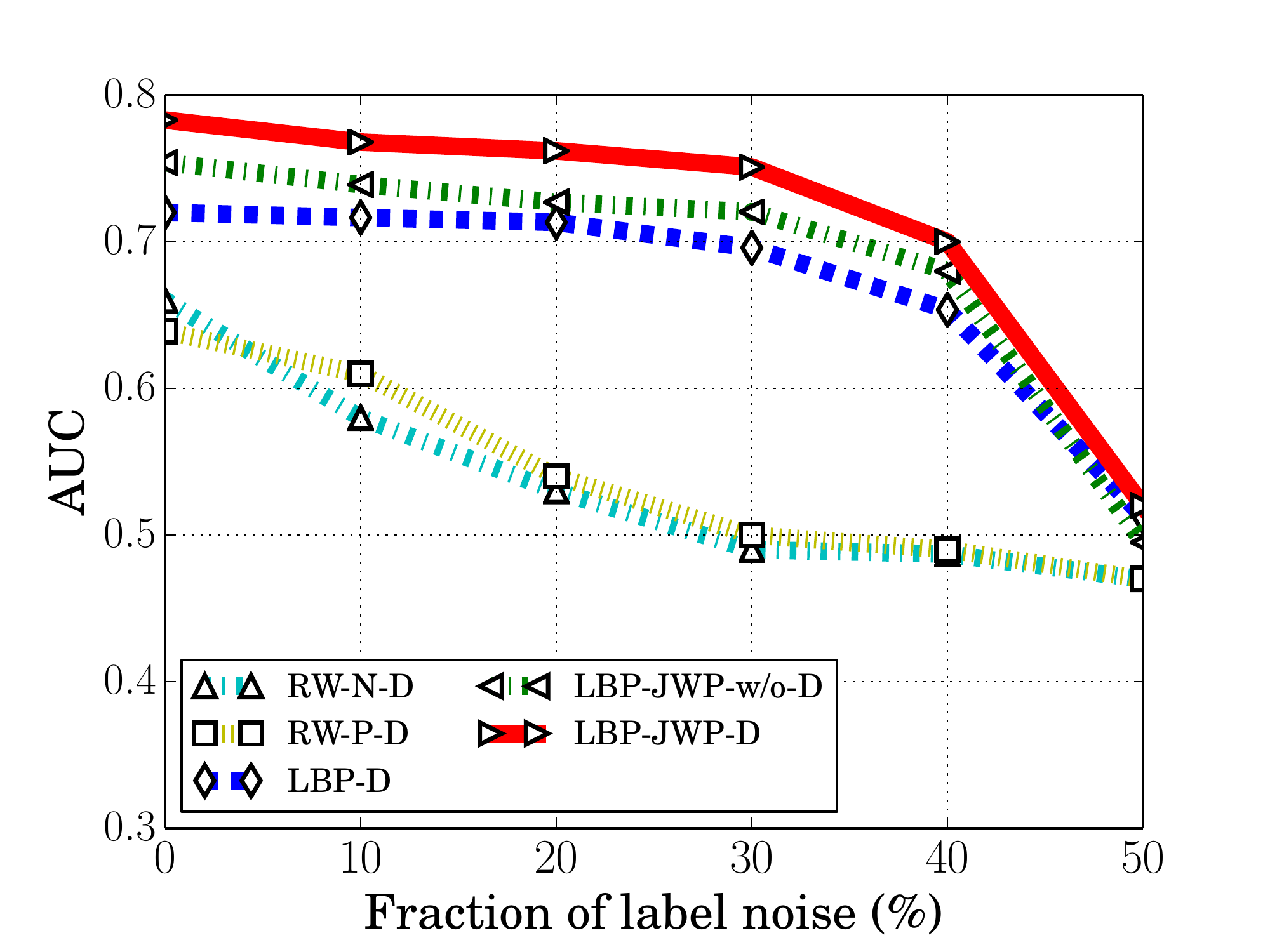} \label{noise-d}}
\caption{Impact of  label noise on Twitter. (a) RW-based methods for undirected graphs; (b) LBP-based methods for undirected graphs; (c) Methods for directed graphs.}
\vspace{-6mm}
\end{figure*}

\subsection{Scalability}

We measure scalability of the compared methods with respect to the number of edges in the graph. 
Since we need graphs with different number of edges, we synthesize graphs using a network generation model. In particular, we use the popular Preferential Attachment model~\cite{Barabasi99} developed by the network science community to generate graphs. 
Specifically, we first use the Preferential Attachment model to synthesize an undirected graph $G$. Then, we treat each undirected edge $(u,v)$ in $G$ as two directional edges $(u,v)$ and $(v,u)$. Finally, we sample 
50\% of directed edges and remove them to construct a directed graph $G'$.  

We run methods for undirected graphs on $G$ and methods for directed graphs on $G'$. 
We note that all the compared methods iteratively solve the posterior reputation scores. 
For fair comparison, we run the iterative processes with the same number of iterations (e.g., 10 in our experiments). 
Figure~\ref{efficiency-u} and Figure~\ref{efficiency-d} respectively show the running time of these methods on undirected graphs and  directed graphs as the number of (undirected and directed) edges increases.  
Note that RW-N-U, RW-P-U, and RW-FLW-U have the same running time and thus we only show RW-P-U for simplicity. 
Similarly, RW-N-D and RW-P-D have the same running time and thus we only show RW-P-D for simplicity.

\myparatight{Our methods have an acceptable computational overhead} We observe that our methods LBP-JWP-U and LBP-JWP-D are less efficient than existing methods. This is because our methods jointly learn edge weights and propagate reputation scores, while existing methods only propagate reputation scores. However, the computational overhead of our methods is acceptable. In particular, 
our methods are 2-3 times slower than state-of-the-art methods.

\myparatight{Undirected graphs vs. directed graphs}  When an undirected graph and a directed graph have the same number of edges, a method for the directed graph is more efficient than its variant for the undirected graph. For instance, RW-P-D is more efficient than RW-P-U, LBP-D is more efficient than LBP-U, LBP-JWP-D is more efficient than LBP-JWP-U. The reason is that RW-P-D traverses each directed edge once, while RW-P-U traverses each undirected edge twice in each iteration of propagating reputation scores; LBP-D (or LBP-JWP-D) only traverses an edge $(u,v)$ twice when the directed edge $(v,u)$ does not exist, while LBP-U (or LBP-JWP-U) traverses each undirected edge twice in each iteration. 

\myparatight{LBP-based methods vs. RW-based methods} For undirected graphs, LBP-U and RW-P-U have the same running time. This is because both methods traverse each undirected edge twice in each iteration of propagating reputation scores.  However, on directed graphs, LBP-D is slightly slower than RW-P-D. This is because RW-P-D traverses each directed edge once in each iteration, while LBP-D traverses an edge $(u,v)$ twice when the directed edge $(v,u)$ does not exist. 
These results are consistent with previous studies~\cite{wang2017sybilscar,wang2017gang}.

\subsection{Robustness to  Label Noise}

We randomly sample $\alpha$\% of labeled positive nodes and labeled negative nodes in the training dataset, and change their labels to be negative and positive, respectively. 
Therefore, the label noise level in the training dataset is $\alpha$\%. 
{We repeat the sampling process 5 times.}
Figure~\ref{noise-u-rw}, Figure~\ref{noise-u-lbp}, and Figure~\ref{noise-d} respectively show the average AUCs on Twitter for undirected graph based methods using RW, undirected graph based methods using LBP, and directed graph based methods, as we increase $\alpha$\% from 0\% to 50\%. 
We cut off at the 50\% because no methods are effective for a label noise level that is higher than 50\%. 

First, LBP-JWP consistently performs the best and can tolerate label noise up to 20\% on the undirected graph and 30\% on the directed graph. For instance,  LBP-JWP-D's AUC only slightly decreases when the label noise is 30\%. Moreover, LBP-JWP has higher AUCs than LBP-JWP-w/o, which shows the effectiveness of our  consistency regularization. 
Second, LBP-based methods and RW-B-U can tolerate relatively larger label noise than RW-N and RW-P.
A possible reason is that LBP-based methods and RW-B-U leverage both labeled positive nodes and labeled negative nodes in the training dataset, while RW-N and RW-P only use labeled negative nodes or labeled positive nodes.

\begin{table}[!tbp]\renewcommand{\arraystretch}{1.8}
\vspace{-2mm}
\centering
\caption{Composition of the manually labeled 1,000 randomly sampled users.}
\label{tab}
{\footnotesize
\begin{tabular}{|c|c|c|c|c|c|}
\hline
\multicolumn{2}{|c|}{\textbf{\small Category}} & \multicolumn{2}{c|}{\textbf{\small LBP-D}} & \multicolumn{2}{c|}{\textbf{\small LBP-JWP-D}} \\ \hline
\multirow{3}{*}{\emph{\bf Sybil}}  &  {\small Suspended Users}  & 41.5\%  & \multirow{3}{*}{92.0\%} &  41.9\% & \multirow{3}{*}{94.9\%} \\ \cline{2-3} \cline{5-5}
                  &   {\small Spammers} &  42.5\% &                   &  44.3\% &                   \\ \cline{2-3} \cline{5-5}
                   &  {\small Compromised Users} & 8.0\%  &                   & 8.7\%  &                   \\ \hline
\multicolumn{2}{|c|}{\emph{\bf Benign}} & \multicolumn{2}{c|}{6.8\%} & \multicolumn{2}{c|}{4.3\%} \\ \hline
\multicolumn{2}{|c|}{\emph{\bf Unknown}} & \multicolumn{2}{c|}{1.2\%} & \multicolumn{2}{c|}{0.8\%} \\ \hline
\end{tabular}
}
\end{table}

\subsection{Case Study on Sina Weibo}

We apply our LBP-JWP-D to detect Sybils on Sina Weibo. 
Specifically, we use the 1,980 labeled users as a training  dataset, leverage LBP-JWP-D to compute a posterior reputation score for every node, and rank all unlabeled nodes in a descending order with respect to their posterior reputation scores. Therefore, the nodes that are ranked higher could be more likely to be Sybils. In practice, social network providers often rely on human workers to manually inspect users and detect Sybils~\cite{sybilrank}. The ranking list produced by our method can be used as a priority list for the human workers, i.e., the human workers can inspect users according to the ranking list. Within the same amount of time, the human workers could find more Sybils when inspecting users according to the ranking list than inspecting users that are randomly picked.

We follow the same strategy of~\cite{wang2017gang} to evaluate the top-ranked 100K users produced by our method. In particular, via manual inspection, Wang et al.~\cite{wang2017gang} classified users on Sina Weibo into five categories: 
\emph{Suspended users}, \emph{Spammers}, \emph{Compromised users}, \emph{Normal users}, and \emph{Unknown users}. 
The first three categories (i.e., suspended users, spammers, and compromised users) are treated as Sybils, and the normal users are treated as benign. We divide the top-ranked 100K users into 10 10K-user intervals. We randomly sample 100 users from each interval, and thus we have 1,000 sampled users in total. We manually inspected the Sina Weibo pages of the 1,000 sampled users and labeled them to be one of the five categories following the same strategy in~\cite{wang2017gang}.

\begin{figure}[!t]
	\centering
	\vspace{-4mm}
	\includegraphics[width=0.42\textwidth]{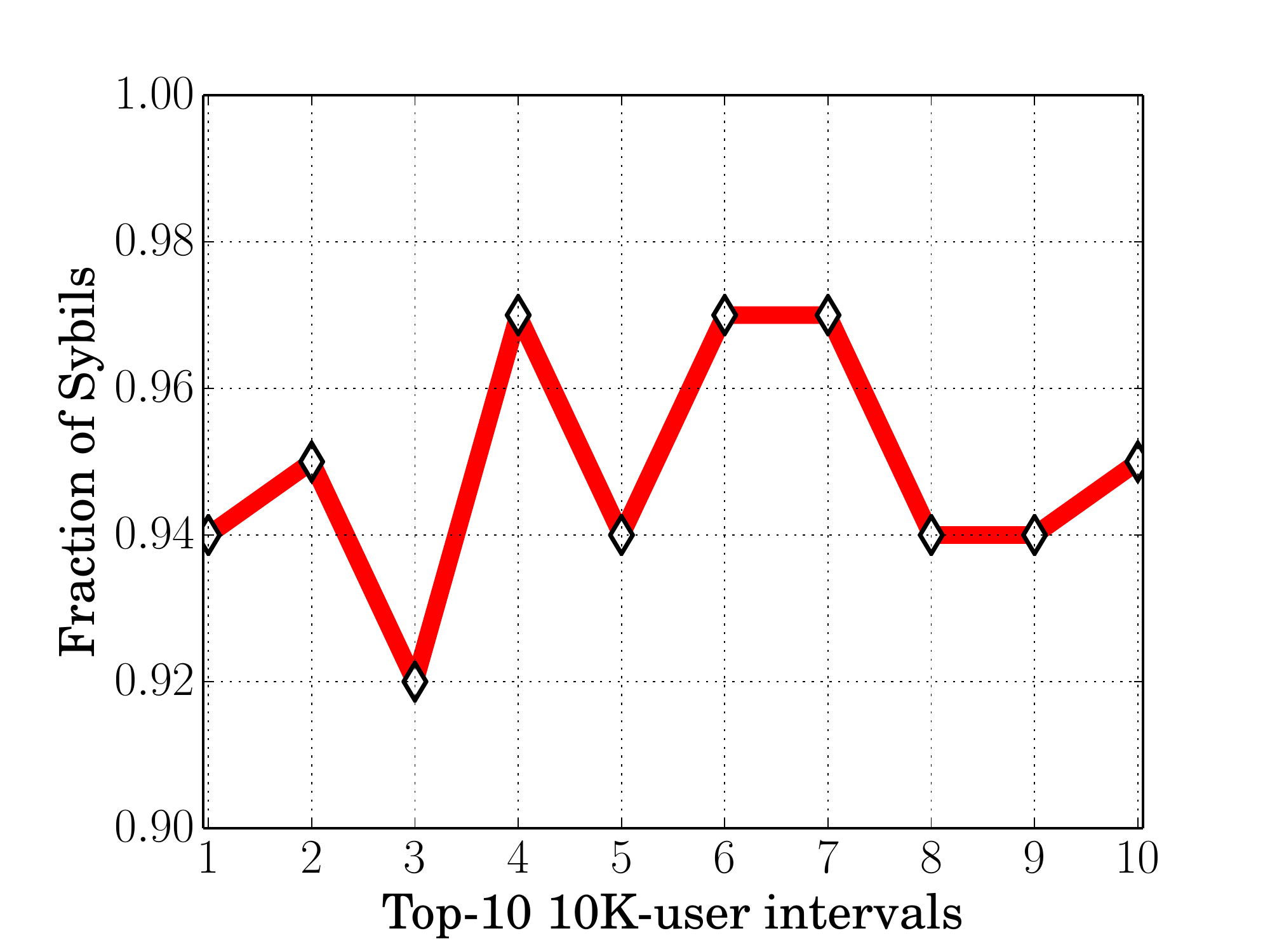}
	\caption{Fraction of Sybils in the sampled 100 users in each of the top-10 10K-user intervals for LBP-JWP-D on the Sina Weibo dataset.} 
	\label{SE-weibo}
	\vspace{-6mm}
\end{figure}

Table~\ref{tab} shows the manual labeling results of the 1,000 sampled users. For comparison, we also include the results for LBP-D, which are from~\cite{wang2017gang} instead of being labeled by us. We note that LBP-D produces a different ranking list from our method, and the sampled 1,000 users are also different for them. 
We observe that 94.9\% of the nodes ranked by our method are Sybils, around 3\% higher than those ranked by LBP-D~\cite{wang2017gang}. We note that \'{I}ntegro~\cite{integro} has a similar accuracy at detecting Sybils in Tuenti. Moreover, our method detects a large number of Sybils that evaded  Sina Weibo's detector. In particular, if a Sybil was detected by Sina Weibo, then the Sybil should have been suspended or deleted. However, 44.3\% of our top-ranked 100K nodes are spammers and 8.7\% of them are compromised users, all of which evaded Sina Weibo's detector. 
Figure~\ref{SE-weibo} shows the fraction of Sybils in the sampled 100 users in each 10K-user interval. 
We observe that even if the training dataset is small (around 0.05\% of all users in the dataset), above 90\% of nodes are Sybils in each of the top-10 10K-user interval ranked by our method.

\section{Discussion and Limitation}
\label{discussion}

\myparatight{Applying our framework to RW-based methods} For conciseness, we focus on applying our framework to LBP-based collective classification methods for graph-based security and privacy analytics. However, our framework is also applicable to RW-based methods. In particular, RW-based methods can also be viewed as iteratively solving system of equations, though the function $f(\mathbf{q}, \mathbf{W}, \mathbf{p})$ is different for RW-based methods. Therefore, our framework can also learn edge weights for RW-based methods. We implemented our framework to learn edge weights for RW-B-U~\cite{jia2017random}, a RW-based method for undirected graphs that incorporates both labeled positive nodes and labeled negative nodes in the training dataset.  
The function $f$ for RW-B-U is  $f(\mathbf{q}, \mathbf{W}, \mathbf{p})=\mathbf{W} \mathbf{p}$, where $w_{uv}$ is the edge weight of $(u,v)$ divided by the weighted degree of $u$. 
Table~\ref{AUC_JWP_RW} shows the AUCs of RW-B-U, RW-JWP-w/o-U, and RW-JWP-U for the four  datasets. RW-JWP-U uses our framework to learn edge weights for RW-B-U, while RW-JWP-w/o-U uses our framework without the consistency regularization to learn edge weights for RW-B-U. Our results show that our framework improves AUCs for RW-B-U and the consistency regularization helps improve AUCs.   

\begin{table}[!tbp]\renewcommand{\arraystretch}{1.8}
\centering
\caption{AUCs of our framework for RW-based methods.}
\label{AUC_JWP_RW}
{\small
\begin{tabular}{|c|c|c|c|c|}
\hline
{\small \textbf{Methods}} & {\small \textbf{Twitter}}  &  {\small \textbf{Sina Weibo}} &   {\small \textbf{Yelp}} &  {\small \textbf{Google+}} \\ \hline
{\small RW-B-U} & {\small 0.63} &   {\small 0.68} &   {\small 0.58} &   {\small 0.63}  \\ \hline
{\small RW-JWP-w/o-U} & {\small 0.66} &   {\small 0.72} &   {\small 0.58} &   {\small 0.65}  \\ \hline
{\small RW-JWP-U} & {\small 0.69} &   {\small 0.74}  &   {\small 0.60} &   {\small 0.68}  \\ \hline
\end{tabular}
}
\end{table}

\myparatight{Incorporating problem-specific characteristics} In this work, our framework only uses the graph structure and can be applied to various security and privacy problems. For a specific security or privacy problem, nodes and edges in the graph could have problem-specific rich features. Our framework can be extended to incorporate such problem-specific features to further enhance detection accuracy. For instance, for Sybil detection in social networks, each node could have rich content and behavior features. Our framework can use these node features to learn the prior reputation scores of nodes. Moreover, an edge could have features such as interaction frequency between the two users. We can model the weight of an edge as a function of the edge's features, e.g., a logistic function. Then, our framework can be used to learn the parameters in the function. Specifically, we solve the optimization problem in Equation~\ref{weightlearning} with respect to such parameters, i.e., we compute the gradient of the objective function with respect to such parameters and then use gradient descent to optimize them.

\myparatight{Security of collective classification methods under adversarial settings} In this work, we evaluate the collective classification methods with respect to accuracy, scalability, and robustness to \emph{random} label noise. It is an interesting future work to study the security of collective classification methods under adversarial settings. For instance, how can an attacker inject carefully crafted label noise to reduce detection accuracy?  

\myparatight{Computational overhead} Theoretically, our method has the same asymptotic time complexity as state-of-the-art random walk based and belief propagation based methods: linear to the number of edges in the graph. Empirically, our method is 2-3 times slower. However, this time overhead is tolerable in practice, especially the targeted security and privacy applications are not time-critical. For instance, on the Twitter dataset with 1.5 billion edges, our method finishes within 3 hours on a server with 512GB memory and 32 cores. Our method can be easily parallelized and should be scalable to graphs with billions of edges on a modern data center.

\myparatight{No manual labels} Like existing methods, our framework relies on a manually labeled training dataset. When there are no such training dataset, our framework may also be applicable. Specifically, Wang et al.~\cite{Wang18RAID} recently proposed SybilBlind, which generalizes LBP-based collective classification methods to the scenarios where no training dataset is available. Their key idea is to  randomly sample some nodes from the graph and treat them as labeled nodes (some of them would be labeled incorrectly, i.e., the training data has noise). Then, they apply a LBP-based method with the sampled training data. Since LBP-based methods are robust to some noise in training data, they repeat the sampling process multiple times and design some method to select the sampling trials that have small label noise. We believe such idea can also be applied to our framework when no manual labels are available, because our framework is also robust to some noise in training data.

\myparatight{Comparing with other graph-based classification methods}
A graph-based classification problem can also be solved via \emph{graph embedding} methods and Graph Convolutional Networks (GCN)~\cite{kipf2017semi}. In particular, a graph embedding method first learns feature vectors for nodes via unsupervised learning, and then learns a classifier (logistic regression classifier in our experiments) based on the feature vectors and the training dataset. We compare with three state-of-the-art graph embedding methods (DeepWalk~\cite{perozzi2014deepwalk}, LINE~\cite{tang2015line}, and node2vec~\cite{grover2016node2vec}) as well as GCN for Sybil detection. We obtained the public source code of these methods from their authors.
 As DeepWalk, node2vec, and GCN are not scalable to the large graphs used in our experiments, we perform experiments on an undirected Facebook graph with synthetic Sybils. The Facebook graph contains 4,039 negative nodes and 88,234 edges. Following prior work~\cite{wang2018structure}, we replicate the nodes and their edges as positive nodes and edges. Moreover, we add 10k edges between the positive nodes and negative nodes uniformly at random. We randomly select 100  positive nodes and 100  negative nodes as the training dataset. A graph embedding method learns a 128-dimension feature vector for each node. We set the number of hidden units of GCN to be 64 and set the dropout rate as 0.5. We observe that our method is more accurate than the compared methods. Specifically, DeepWalk, LINE, node2vec, and GCN achieve AUCs of 0.62, 0.94, 0.74, and 0.88, respectively, while our LBP-JWP-U achieves an AUC of 1. We speculate the reason is that graph embedding aims to learn general feature representations that could be used for various purposes (classification is just one of the purposes) and GCN is not able to learn discriminative hidden features for classification.

\section{Conclusion and Future Work}
\label{conclusion}

In this work, we propose a new framework to learn edge weights for graph-based security and privacy analytics. Our framework can be applied to various security and privacy problems including, but not limited to, Sybil detection in social networks, malware detection, fake review detection, and attribute inference attacks.  
Our framework can be incorporated into a state-of-the-art collective classification method (both RW-based methods and LBP-based methods) to further enhance its classification accuracy with an acceptable computational overhead. Our results demonstrate that jointly learning edge weights and propagating reputation scores is effective for graph-based security and privacy analytics.  

Interesting future work includes 1) incorporating problem-specific characteristics into our framework to further enhance accuracy for a specific security or privacy problem, and 2) studying security of collective classification methods under adversarial settings.

\section*{Acknowledgements}
We thank the anonymous reviewers for their constructive comments. 
This work is supported by the National Science Foundation under
Grants No. CNS-1750198 and a research gift from JD.com. 
Any opinions, findings and conclusions or recommendations expressed in this material are those of the author(s) and do not necessarily reflect the views of the funding agencies.

\balance
{
\bibliographystyle{IEEEtran}
\bibliography{refs}
}

\end{document}